\documentclass{iau} 
\usepackage{graphicx}
\usepackage{natbib}
\usepackage{enumitem}

\pubyear{2022}
\volume{370}
\jname{Winds of Stars and Exoplanets}
\editors{A.~A. Vidotto, L.~Fossati \& J.~Vink, eds.}

\newcommand{\beq}{\begin{equation}}
\newcommand{\eeq}{\end{equation}}
\newcommand{\beqa}{\begin{eqnarray}}
\newcommand{\eeqa}{\end{eqnarray}}

\newcommand{\apj}{$ApJ$}
\newcommand{\aap}{$A\&A$}

\newcommand{\Mstar}{\ensuremath{M_{\ast}}}
\newcommand{\Rstar}{\ensuremath{R_{\ast}}}
\newcommand{\estar}{\ensuremath{\eta_{\ast}}}

\newcommand{\vinf}{\ensuremath{v_{\infty}}}
\newcommand{\Ralf}{\ensuremath{R_{\rm A}}}
\newcommand{\Rkep}{\ensuremath{R_{\rm K}}}

\newcommand{\etas}{\eta_\ast}
\newcommand{\ra}{R_\mathrm{A}}
\newcommand{\rk}{R_\mathrm{K}}
\newcommand{\rs}{R_\ast}
\newcommand{\teff}{T_\mathrm{eff}}
\newcommand{\Beq}{B_\mathrm{eq}}
\newcommand{\bp}{B_\mathrm{p}}
\newcommand{\mdot}{\dot{M}}

\title[Winds and Magnetospheres] 
{Winds and magnetospheres  from stars and planets: similarities and differences}

\author[Stan Owocki]   
{Stan Owocki
}

\affiliation{Department of Physics \& Astronomy, Bartol Research Institute, 
University of Delaware, Newark, DE 19716 USA \\ email: {\tt owocki@udel.edu}}

\begin{document}

\maketitle

\begin{abstract}
Both stars and planets can lose mass through an expansive wind outflow, often constrained or channeled by magnetic fields that form a surrounding magnetosphere. The very strong winds of massive stars are understood to be driven by line-scattering of the star's radiative momentum, while in the Sun and even lower-mass stars a much weaker mass loss arises from the thermal expansion of a mechanically heated corona. In exoplanets around such low-mass stars, the radiative heating and wind interaction can lead to thermal expansion or mechanical ablation of their atmospheres. Stellar magnetospheres result from the internal trapping of the wind outflow, while planetary magnetospheres are typically shaped by the external impact from the star's wind. But in both cases the stressing can drive magnetic reconnection that results in observable signatures such as X-ray flares and radio outbursts. This review will aim to give an overview of the underlying physics of these processes with emphasis on their similarities and distinctions for stars vs. planets.
\end{abstract}

\firstsection 
\section{Introduction}

\begin{figure}
\begin{center}
\includegraphics[width=4.in]{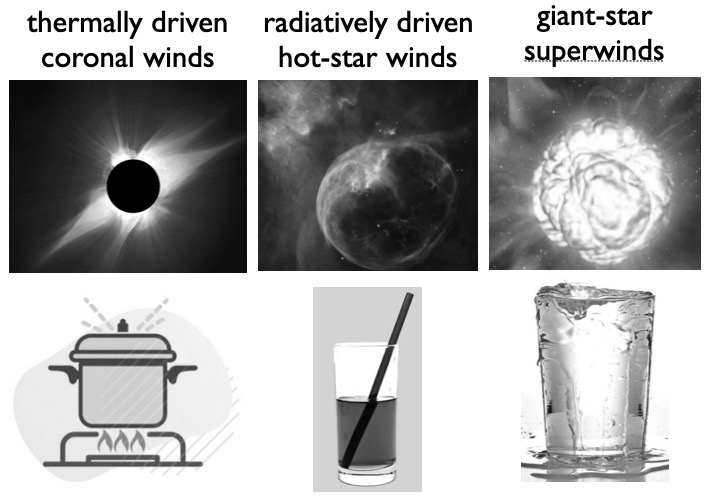} 
 \caption{Icons to represent analogies for processes inducing the three different types of steady stellar wind outflow.
 }
\label{fig01}
\end{center}
\end{figure}

To set the stage for this symposium's exploration of ``The Winds from Stars and Exoplanets", I have been asked to review the similarities and 
differences in the physical processes that drive wind outflows from stars vs. planets, including the distinct roles that magnetic fields play in trapping
and diverting plasma flows in an associated {\em magnetosphere}.

As illlustrated in figure \ref{fig01}, one can identify three broad classes of stellar wind:
\begin{enumerate}[label=\arabic*.~]
	\item{} the pressure-driven coronal wind of the Sun and other cool stars;
	\item{} radiatively driven winds from OB stars;
	\item{}  the slow ``overflow'' mass loss from highly evolved giant stars.
\end{enumerate}
For the last a key is the greatly reduced gravity, which allows even surface convection or pulsations to eject outer layers to escape,
somewhat like Roche-lobe overflow in binary systems.
The resulting mass loss can be irregular and difficult to quantify, and since it has little  overlap with winds from planets,
we set this aside to focus  on the former two.

We first review (\S 2) the radiative driving  of OB winds, 
which occurs through a kind of line-driven ``suction" effect. 
We then discuss (\S 3) how the winds of the Sun and other cool stars are, in contrast, driven by the gas pressure associated with hot corona, with mass loss representing now an escape valve analogous to that of pressure cooker.

This provides a basis for discussion (\S 4) of planetary outflows, which are likewise largely driven by gas-pressure expansion, now powered by the UV and X-ray
heating from the chromospheric and coronal emission from the underlying cool star.

We conclude with a review (\S 5) of stellar magnetospheres, contrasting their inside-out, internal filling by the stellar wind with the outside-in, external 
stress imposed on planetary magnetospheres by the  wind of their host star.

\section{ Radiatively driven winds from OB stars}

\subsection{Radiative acceleration and Eddington parameter}

In hot stars with a high luminosity, the outward force from scattering of stellar radiation can overcome gravity and so drive a 
stellar wind outflow.
For opacity $\kappa_\nu$ at a frequency $\nu$ with radiative flux $F_\nu$, the total radiative acceleration depends on the frequency integral,
\beq
g_{rad} = \int_0^\infty d\nu \, \frac{\kappa_\nu F_\nu}{c} \equiv \frac{{\bar \kappa}_F F}{c}
\,
\label{eq:grad}
\eeq
where the last equality defines the flux-weighted mean opacity, ${\bar \kappa}_F$, 
with $F$ the bolometric flux.

In the idealized case of continuum scattering by free electrons, 
${\bar \kappa}_F$ 
 is just equal to the electron scattering opacity $ \kappa_e = (1+X) 0.2 = 0.34  \, {\rm cm}^2$/g, where the latter value applies to  a fully ionized gas 
 with solar  hydrogen mass fraction $X=0.72$.
The ratio of the associated radiative acceleration to gravity defines the classical Eddington parameter,
\beq
\Gamma_e \equiv \frac{\kappa_e F/c}{g} = \frac{\kappa_e L}{4 \pi GMc}  = 2.6 \times 10^{-5} \frac{L/L_\odot}{M/M_\odot}
\sim 0.26 \left ( \frac{M}{100 \, M_\odot} \right )^2
\, ,
\label{eq:game}
\eeq
wherein the inverse-square radial dependence of both the radiative flux $F=L/4\pi r^2$ and gravity $g=GM/r^2$ cancels, 
showing this Eddington parameter depends only  on the ratio $L/M$ of luminosity to mass.
The third equality shows that $\Gamma_e$  is very small for stars like the Sun; 
but if one applies the standard main-sequence mass-luminosity scaling $L \sim M^3$, the last equality shows that 
 massive stars can have Eddington parameters that approach unity.
This provides a basic rationale for the upper limit to stellar mass,  which is empirically found to be around $200 M_\odot$,
remarkably  close to the mass for which (\ref{eq:game}) gives $\Gamma_e \approx 1$.
Stars that approach or exceed this classical Eddington limit  can have strong eruptive mass loss, as thought to occur 
in eruptive Luminous Blue Variable stars like $\eta$\,Carinae.

But since generally ${\bar \kappa}_F \gg \kappa_e$, even stars with $\Gamma_e \ll 1$ can have a total $\Gamma > 1$, and 
drive a more steady-state  {\em stellar wind} mass loss.
Ignoring the small gas-pressure acceleration, the associated steady-state acceleration has the scaling,
\beq
v \frac{dv}{dr} = - \frac{GM}{r^2} +  \frac{\kappa L}{4 \pi r^2} = (\Gamma - 1) \frac{GM}{r^2}
\, ,
\label{eq:eom1}
\eeq
where to simplify the notation, we have set ${\bar \kappa}_F = \kappa$.
For constant $\Gamma$, this can be trivially integrated to give the variation of wind velocity with radius $r$,
\beq
v(r) = v_\infty \, (1- R/r)^{1/2} ~~ ; ~~ v_\infty = v_{esc} \sqrt{\Gamma - 1}
\, ,
\label{eq:vinf}
\eeq
which shows that  the wind terminal speed $v_\infty$ just scales with the escape speed $v_{esc} \equiv \sqrt{2 GM/R}$ from the 
wind initiation surface radius $R$.
We thus see that $\Gamma$ represents an effective {\em anti-gravity}, which 
for the simple case $\Gamma=2$ gives a direct gravitational {\em reversal}, with material flying away from the star,
asymptotically reaching the escape speed.

Moreover, if we multiply (\ref{eq:eom1}) by $4 \pi \rho r^2 dr$, and note the standard definition of mass loss rate ${\dot M} \equiv 4 \rho v r^2$,
 we find upon integration a relationship between the 
wind momentum ${\dot M} v_\infty$ and its optical depth $\tau \equiv \int_R^\infty \kappa \rho dr$,
\beq
{\dot M} v_\infty = \frac{\tau L}{c} \left ( \frac{\Gamma -1 }{\Gamma} \right )
\, .
\label{eq:momtau}
\eeq
OB stars generally have $\tau < 1$ and so fall within the single-scattering limit, ${\dot M} v_\infty < L/c$.
In contrast, Wolf-Rayet (WR) stars can have $\tau \approx 1 - 10$ and so require multi-line scattering to explain their
optically thick winds with ${\dot M} v_\infty > L/c$.

\begin{figure}
\begin{center}
\includegraphics[width=5.0in]{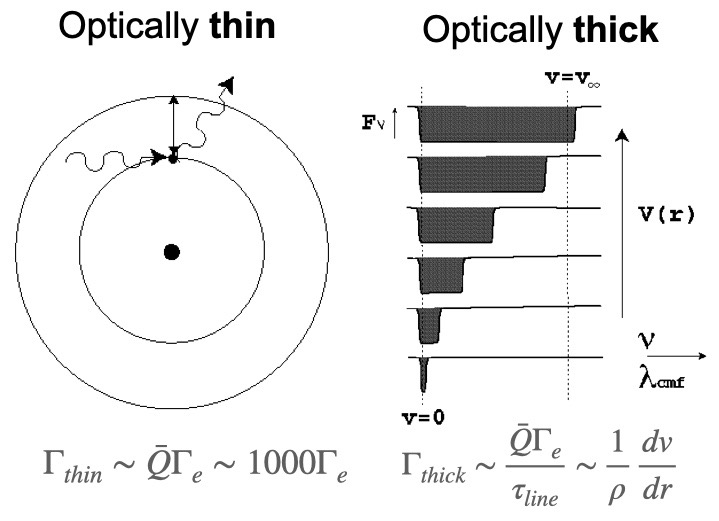} 
\vspace{-0.1in} 
 \caption{
 {\em Left}: Illustration of the resonance nature of line opacity for an optically thin line,
 for which the spectral average over the resonance quality (${\bar Q}$) results in an optically 
 thin line acceleration that is of order a thousand times the gravitationally scaled acceleration from 
 electron scattering  $\Gamma_e$.
 {\em Right}: The corresponding acceleration for an optically thick line, which is reduced by the line optical thickness $\tau_{line}$,
 which in an accelerating wind scales with the ratio of local density to velocity gradient.
This line desaturation is a consequence of the Doppler shift of the line absorption from the increasing wind velocity $v$,
resulting in a net line force that scales as $\Gamma_{thick} \sim (1/\rho) (dv/dr)$.
}
\label{fig02}
\end{center}
\end{figure}

\subsection{The CAK model for line-driven stellar winds}

In practice, the enhancement of the flux-weighted mean opacity above the simple electron scattering value
results mainly from the {\em bound-bound resonance} of radiation with electrons bound into heavy ions ranging from CNO to Fe and Ni.
As illustrated in the left panel of figure \ref{fig02}, the resonance nature of such bound-bound line-scattering greatly enhances the opacity,
typically by a factor ${\bar Q} \gtrsim 10^3$ \citep{Gayley95}.
This means any star with $\Gamma_e \gtrsim 1/{\bar Q} \approx 10^{-3} $ can have a line-force that overcomes gravity and so drive a wind outflow.

In practice this maximal line acceleration from optically {\em thin} scattering is reduced by the saturation of the reduced flux within an optically
 {\em thick} line.
But as  illustrated in the right panel of figure \ref{fig02},  the Doppler shift associated with the wind acceleration acts to {\em desaturate}
this line absorption, effectively sweeping the absorption through a broad frequency band, extending out to the frequency associated with the Doppler shift from the wind terminal speed, $v_\infty$.
This wind Doppler shift of line resonance concentrates the interaction of continuum photons
into a narrow resonance layer with width set by the Sobolev length,  $\ell \equiv v_{th}/(dv/dr)$ 
\citep{Sobolev60}, associated with 
acceleration through the ion thermal speed $v_{th}$ that broadens the line profile.
In the outer wind where $\ell \ll r$, the line acceleration for optically thick lines is reduced by $1/\tau$, where the Sobolev optical depth $\tau \equiv {\bar  Q}  \kappa_e \rho \ell$, giving then a line acceleration $\Gamma_{thick} \sim (1/\rho) (dv/dr)$ that itself scales with the wind acceleration.

Within this Sobolev approximation, \citet[][hereafter CAK)]{CAK75} developed a formalism that accounts for
the  cumulative radiative acceleration from a {\em power-law ensemble} of both optically thick and thin lines, 
\beq
\Gamma_{CAK} \approx \frac{{\bar Q} \Gamma_e}{({\bar Q}t )^\alpha} \gtrsim 1 ~~ ; ~~ t \equiv \kappa_e c \frac{\rho}{dv/dr}
\, ,
\label{eq:gamcak}
\eeq
where the CAK power index $\alpha \approx 0.5 - 0.7$ characterizes the relative number of strong vs. weak lines.
Applying (\ref{eq:gamcak}) into the equation of motion (\ref{eq:eom1}) and using the fact that the critical solution requires $v(dv/dr) \sim GM/r^2$,
we  find the maximal CAK mass loss rate has the scaling
\beq
{\dot M} \sim \frac{L}{c^2} \left ( \frac{{\bar Q} \Gamma_e}{1-\Gamma_e} \right )^{-1+1/\alpha}
\, .
\label{eq:mdotcak}
\eeq
For canonical values $\alpha=1/2$, ${\bar Q} = 2000$, and $L= 10^5 L_\odot$, this CAK scaling gives ${\dot M} \approx 10^{-5} M_\odot$/yr,
which is indeed a billion times the mass loss rate of the solar wind!
Much as in the solar wind, the terminal wind speeds scale with the surface escape speed $v_{esc} \equiv \sqrt{2GM/R}$, with values up to $v_\infty \approx 2000$\,km/s.

An overall point is that in such models the onset of line-driving near the sonic point represents an effective {\em line-driven suction}, which draws up mass from the underlying hydrostatic equilibrium of the subsonic region.
The reduction in pressure from the outer line-driving induces the underlying subsonic region to expand upward, much as
 the suction on a straw draws up liquid from a glass  (see figure \ref{fig01}).
 This outside-in suction contrasts with the inside-out thermal expansion of a pressure cooker, and of the analogous gas pressure-driven solar wind
 discussed below in \S 3.
 
 \subsection{The Conti mechanism}

A longstanding question is whether such line-driven winds might, over an O-star's main sequence lifetime, lead to loss of the star's Hydrogen envelope,
representing the so-called ``Conti mechanism" for producing the strong depletion of Hydrogen inferred for Wolf-Rayet (WR) stars.
Using the scaling $L \sim M^3$ to estimate the star's main sequence lifetime $t_{\rm ms} \approx 1 \, {\rm Myr}  \, (100 M_\odot/M)^2$,
we find for the canonical line opacity factor ${\bar Q} = 2000$ that the cumulative mass loss fraction follows the scalings,
\beqa
\frac{{\dot M} t_{\rm ms}}{M} &\approx &~ 0.18 \left ( \frac{M}{100 M_\odot} \right )^2 ~ ; ~ \alpha = 1/2
\nonumber
\\
&\approx & 0.016 \left ( \frac{M}{100 M_\odot} \right )^1 ~ ; ~ \alpha = 2/3
\label{eq:conti}
\eeqa
which shows a very sensitive dependence on the CAK power index $\alpha$.
When corrected for wind clumping, empirically inferred mass loss rates agree better with predictions for higher $\alpha \approx 0.6$, 
and so seem to disfavor the Conti mechanism for H-envelope stripping.

As discussed in the review by A. Sander in these proceedings, quantitative models of OB-wind mass loss endeavor to derive the line-driving opacity self-consistently from NLTE solution of excitation and ionization of key driving ions.

\begin{figure}[b]
\begin{center}
\includegraphics[width=2.6in]{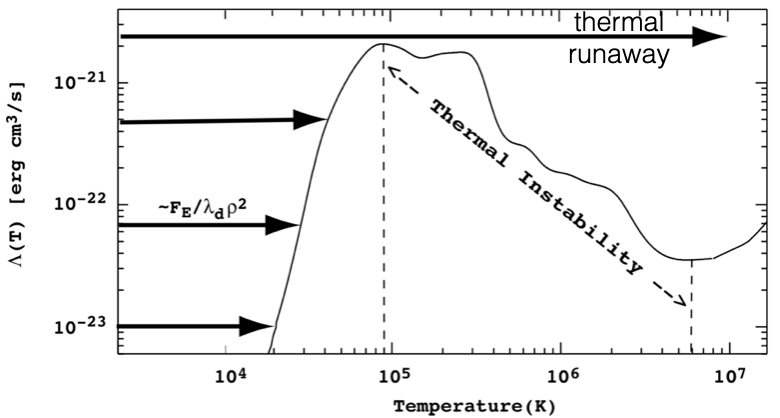} 
\includegraphics[width=2.6in]{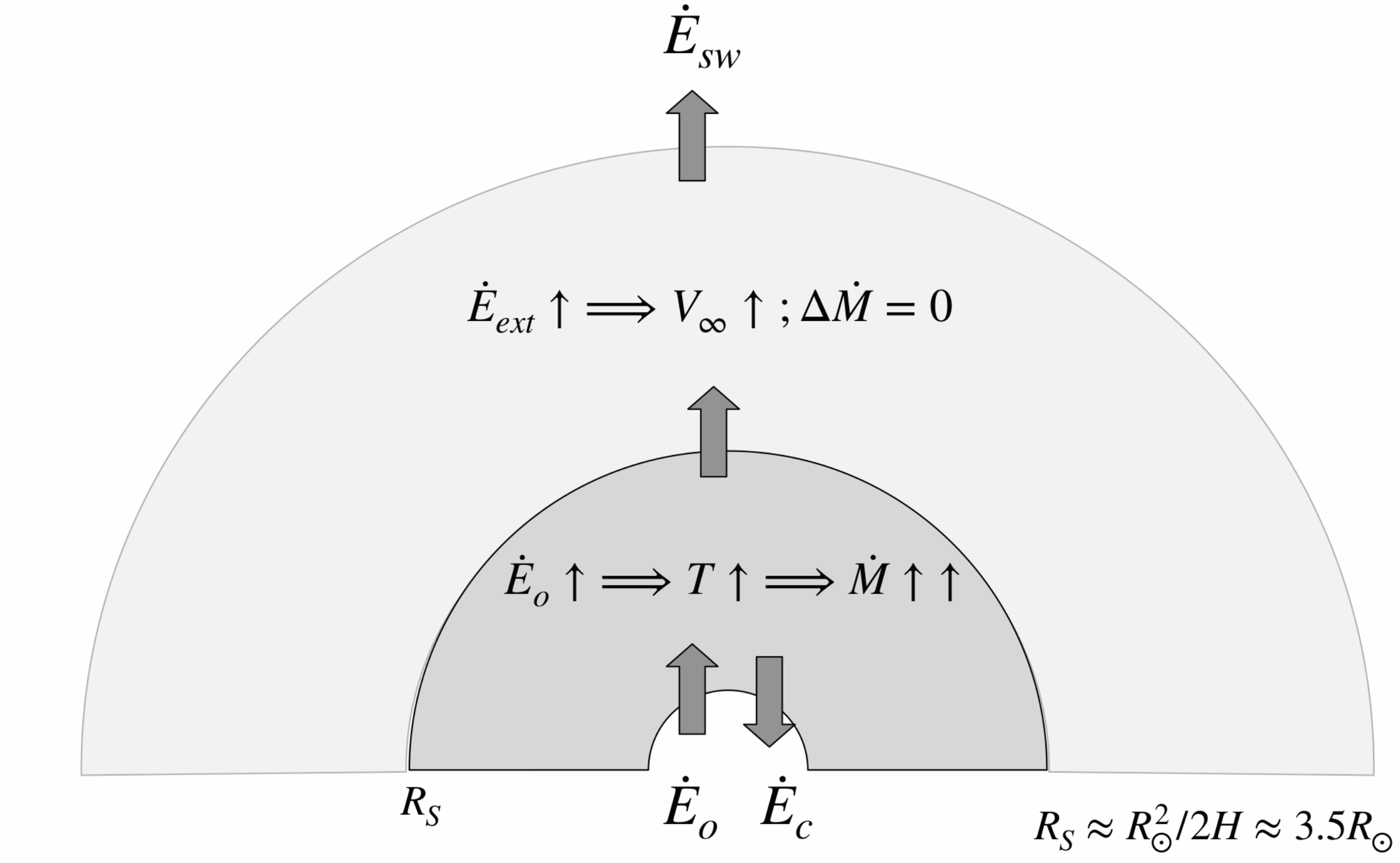} 
 \caption{Left: Illustration of mechanical energy flux $F_E$ deposited over a damping is balance optically thin cooling $ n_e n_i \Lambda(T)$;
the exponential decrease in densit $\rho$ with height leads to high temperature and eventual thermal run above the peak in cooling function $\Lambda$.  Right: Schematic to illustrate energy input into the nearly hydrostatic, subsonic coronal base vs. that into the supersonic wind
 above the Parker sonic radius $R_s$.
 For the former, the net input vs. loss by conduction back into the underlying atmosphere leads to coronal heating that sets
 the coronal temperature and location and density at the sonic radius, thus fixing the associated wind mass loss rate.
 For the latter, any further energy addition increases the wind flow speed, with the wind expansion providing the primary mechanism
 to carry out the total net amount of coronal heating.}
   \label{fig03}
\end{center}
\end{figure}

\section{Pressure-driven coronal winds}

\subsection{Runaway heating of the solar corona and its pressure extension}

In the Sun and other cool stars with surface temperatures below  $10^4$\,K, the recombination of Hydrogen leads to strong subsurface convection
and an associated magnetic turbulence that drives mechanical heating of the upper atmosphere.
As illustrated in the left panel of figure \ref{fig03}, deposition of mechanical energy flux $F_E$ over a damping length $\lambda_d$
must be balanced by radiative cooling. In the upper atmosphere this requires excitation of ions by electrons, so the associated cooling
rate per unit volume scales with density-squared, $\rho^2 \sim n_e n_p $, multiplied by an optically thin cooling function $\Lambda (T)$
that reflects the excitation and ionization of the emitting ions.

As the density decreases exponentially with height, balancing the local heating $F_E/\lambda$ occurs at progressively higher temperatures,
reflecting the initially steep increase in $\Lambda (T)$ from more energetic collisional excitation.
However, for temperatures $T \gtrsim 10^5$\,K, such collisions begin to ionize away the bound electrons, reducing the efficiency of radiative
cooling, and causing the $\Lambda (T)$ to decline with increasing $T$.
Since radiative cooling can no longer balance the heating, this leads to a {\em thermal runaway} to coronal temperatures $T>$MK, 
limited now by inward thermal conduction from the corona to underlying atmosphere, as illustrated in right panel of figure \ref{fig03}.

Maintaining hydrostatic equilibrium at the coronal base radius $R$ implies a pressure scale height,
\beq
H \equiv \frac{P}{|dP/dr|} = 
\frac{k T}{\mu g} = 
\frac{2 a^2}{v_{esc}^2 }R  \equiv
\frac{T}{T_{esc}} \, R
\, ,
\label{eq:hodef}
\eeq
where  $g \equiv GM/R^2$ and $v_{esc} \equiv \sqrt{2GM/R}$ are the surface gravity and escape speed,
and for a gas with molecular weight $\mu$, the isothermal sound scales with temperature as $a \equiv \sqrt{k T/\mu}$.
The last equality defines an ``escape temperature'' at which $H=R$, which for solar parameters has a value $T_{esc} \approx 13.8$\,MK.

In the solar photosphere the low temperature  $T \ll T_{esc}$ implies $H \ll R$ and thus a fixed gravity $g$,  with pressure declining exponentially 
with scale height $H$.
But in the corona, the higher temperature  gives $ T/T_{esc} = H/R \lesssim 1$, implying one now has to account for the radial decline in gravity in the 
hydrostatic balance. 

For a simple isothermal model, the pressure stratification now takes the form,
\beq
P(r) = P(R)  e^{(R/H)(R/r-1)}
\, .
\label{eq:pcor}
\eeq
In contrast to the exponential decline in the photosphere,  the coronal pressure at large radii now asymptotically approaches a finite value, $P_\infty$ .
Relative to the initial pressure $P_o \equiv P(R)$ at the coronal base, the total drop in pressure for a hydrostatic, isothermal corona is given by
\beq
\frac{P_o}{P_\infty} \approx e^{13.8 {\rm MK}/T} ~~ ; ~~ \log \frac{P_o}{P_\infty} \approx \frac{6}{T/{\rm MK}}
\, .
\label{eq:pobpinf}
\eeq
The latter equality shows the pressure drops by 6 decades for $T=1$\,MK, and only 3 decades for $T=2$\,MK.

By comparison, from the solar transition region at the coronal base to the interstellar medium, the 
pressure drop is  actually much greater, $\log (P_{tr}/P_{ism}) \approx 12$.
The upshot is that an extended, hot corona can {\em not} be maintained in hydrostatic equilibrium;
instead, as shown by the outward streamers from the eclipse image in figure \ref{fig03}, it must undergo an outward, supersonic 
{\em expansion} known as the solar wind.

As illustrated by the right panel of figure \ref{fig03}, this solar wind expansion can be thought of as analogous to the release valve of a pressure cooker, driven fundamentally by mechanical heating generated by magnetic turbulence in the underlying solar atmosphere.
Some of this upward energy flux is lost back to the solar atmosphere through thermal conduction,
but the net effect leads to a thermal runaway that raises the coronal temperature to temperature within a factor ten of $T_{esc}$.

\begin{figure}[b]
\begin{center}
\includegraphics[width=2.6in]{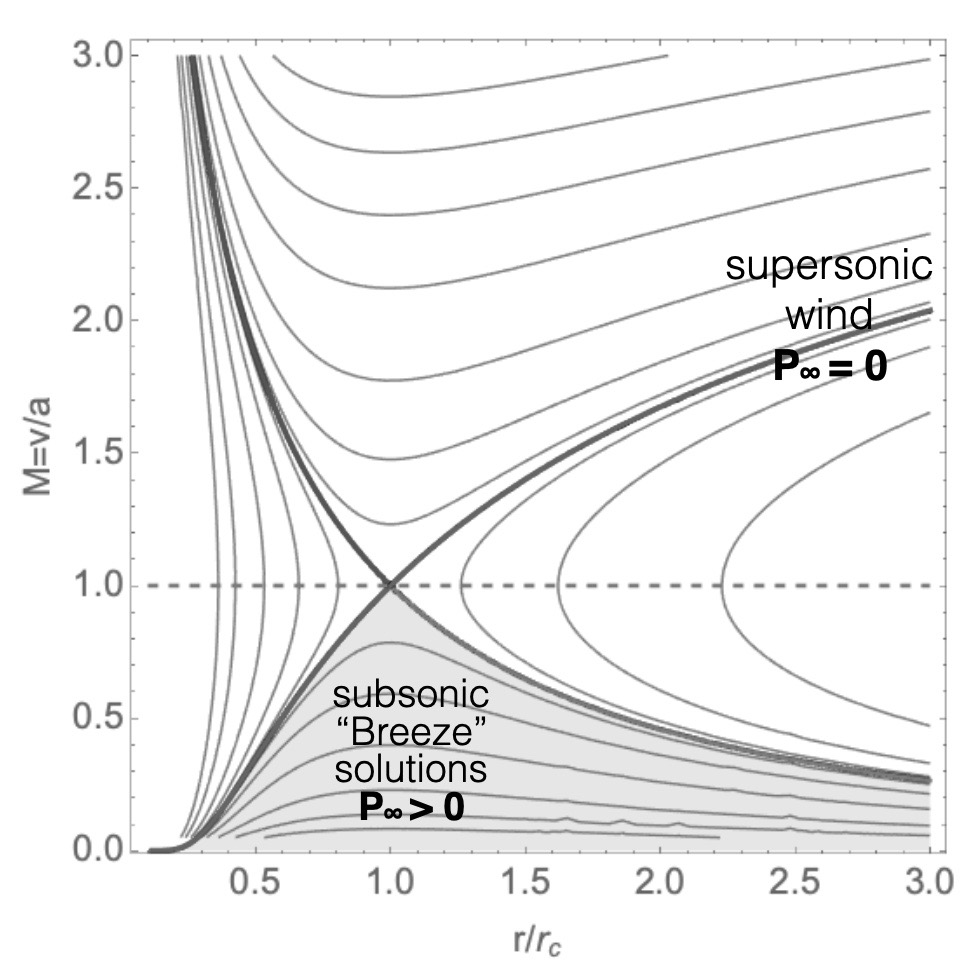} 
\includegraphics[width=2.6in]{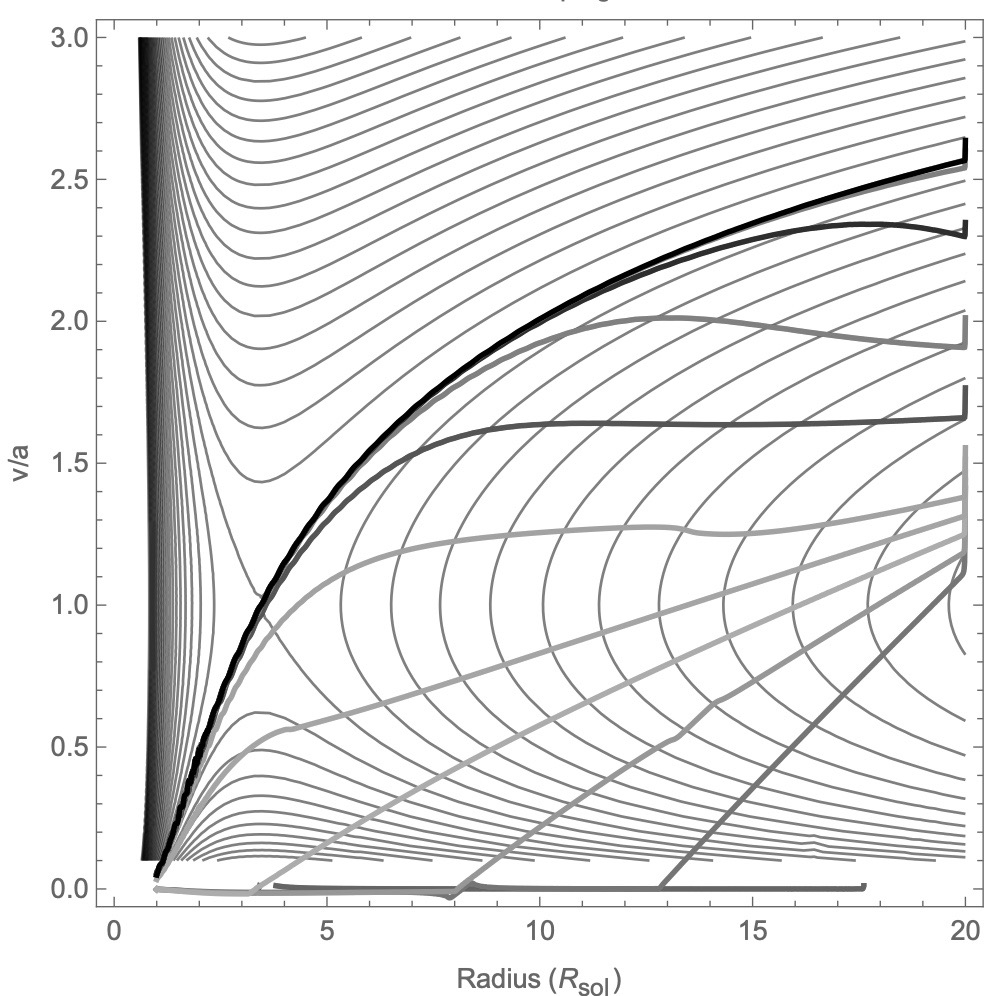}
 \caption{
 Left: Solution topology for Mach number $v/a$ vs. scaled radius $r/r_c$ for an isothermal corona.
 Solutions with initially low-speed at the wind base include subsonic ``breeze'' solutions, which however again have a large 
 terminal pressure, $P_\infty$.  The transonic wind solution is the only with low enough $P_\infty$ match the low pressure of
 the interstellar medium.
 Right: Steady-state velocity topology overplotted with snapshots at $\Delta t = 20$\,ks intervals of a time-depedent
 hydrodynamical simulations for evolution away from an initially hydrostatic corona.
 As the high pressure reacts to the lower pressure outer boundary, an expansion develops that eventually evolves 
 to the supersonic solar wind solution. }
   \label{fig04}
\end{center}
\end{figure}

\subsection{Isothermal Solar Wind}

The imbalance between outward pressure and inward gravity gives rise to an outward acceleration, which for a steady-state
outflow takes the form
\beq
v \, \frac{dv}{dr} = - \frac{GM}{r^2} - \frac{1}{\rho} \frac{dP}{dr}
\, .
\label{eq:sseom}
\eeq
For an isothermal case with $P=\rho a^2$, one can use mass conservation of the spherical outflow, $\rho v r^2 = {\dot M}/4 \pi$=constant,
to eliminate the density $\rho$ in favor of the radius $r$ and flow speed $v$.
This allows one to split the pressure gradient force into terms that scale with the velocity and the sphericity,  giving
\beq
\left (v - \frac{a^2}{v} \right ) \frac{dv}{dr} = 
-\frac{GM}{r^2} + \frac{2 a^2}{r}
\, .
\label{eq:solwineom}
\eeq
In the subsonic region $v \ll a$, this reduces to the condition for hydrostatic equilibrium.

But at a critical (a.k.a. ``Parker") radius, 
\beq
r_c
= \frac{GM}{2a^2} 
= \frac{R^2}{2H} 
=   R \left ( \frac{T_{esc}}{2T} \right )
\, ,
\label{eq:Rs}
\eeq
the spatial component of pressure balances gravity, so making the RHS vanish.
The LHS can likewise vanish if either $dv/dr=0$ or $v=a$.

Figure \ref{fig04} plots Mach number $M \equiv v/a$ vs. scaled radius $r/r_c$ for the  overall topology of solutions. 
We can eliminate all the bi-valued solutions, as well as those that start at supersonic speeds near the coronal base.
There is a class of subsonic solutions that have a velocity peak, with thus $dv/dr=0$, at the critical radius $r=r_c$;
but, like the hydrostatic case, all such ``breeze" solutions asymptote to a finite pressure at large radii, and thus again
cannot match the required low pressure of the interstellar medium.

The only solution that matches both the requirement of low, subsonic speed at the coronal base $v(R) \ll a$,
 and vanishing pressure at large distances, $P_\infty \rightarrow 0$, is the critical transonic solution with $v(r_c)=a$.

\subsection{Energy balance and mass loss rate}

While such isothermal  models provide a basic rationale for the supersonic outflow of the solar wind, they effectively ignore
the central physics of what energy mechanisms keep the corona hot against the adiabatic cooling of the wind expansion.
Moreover, because the density  scales out of the equation of motion (\ref{eq:solwineom}), they do not  provide any basis for 
determining the associated wind mass loss rate, ${\dot M} \equiv 4 \pi \rho v r^2$.
For that one must consider the overall wind {\em energy} balance, which we cast here in terms of integration from some stellar base radius $R$
to an outer radius $r$,
\beq
{\dot M} \left [ \frac{v^2}{2} - \frac{GM}{r} + \frac{5}{2} \frac{P}{\rho} \right ]_R^r = 4\pi \int_R^r  r'^2 q_{net} \, dr' + 4 \pi [ R^2 F_{c*} - r^2 F_c (r) ]
\, .
\label{eq:inten}
\eeq
The terms in parenthesis on the LHS represent the wind kinetic energy,  potential energy, and gas enthalpy,
while the RHS include the integral of the local net heating per unit volume, and the net difference in conductive heat flux.

Since the change in gas enthalpy is generally relatively small, we can write the energy balance over the full wind by
\beq
{\dot M} \left ( \frac{v_\infty^2}{2} + \frac{v_{esc}^2}{2} \right ) \approx Q_{heat} - Q_{rad} - Q_{cond}
\, ,
\label{eq:toten}
\eeq
where $Q_{rad}$ and $Q_{cond}$ are the energy losses from radiation and conduction, and the total heating $Q_{heat}$ is associated with MHD waves and/or magnetic reconnection.
Since the wind terminal speed is generally comparable to the escape speed, $v_\infty \approx v_{esc}$,
we see that the mass loss rate scales directly with the net heating vs. cooling,
\beq
{\dot M} \approx \frac{Q_{net}}{v_{esc}^2}
\, .
\label{eq:mdotq}
\eeq

\begin{figure}[b]
\begin{center}
\includegraphics[width=5in]{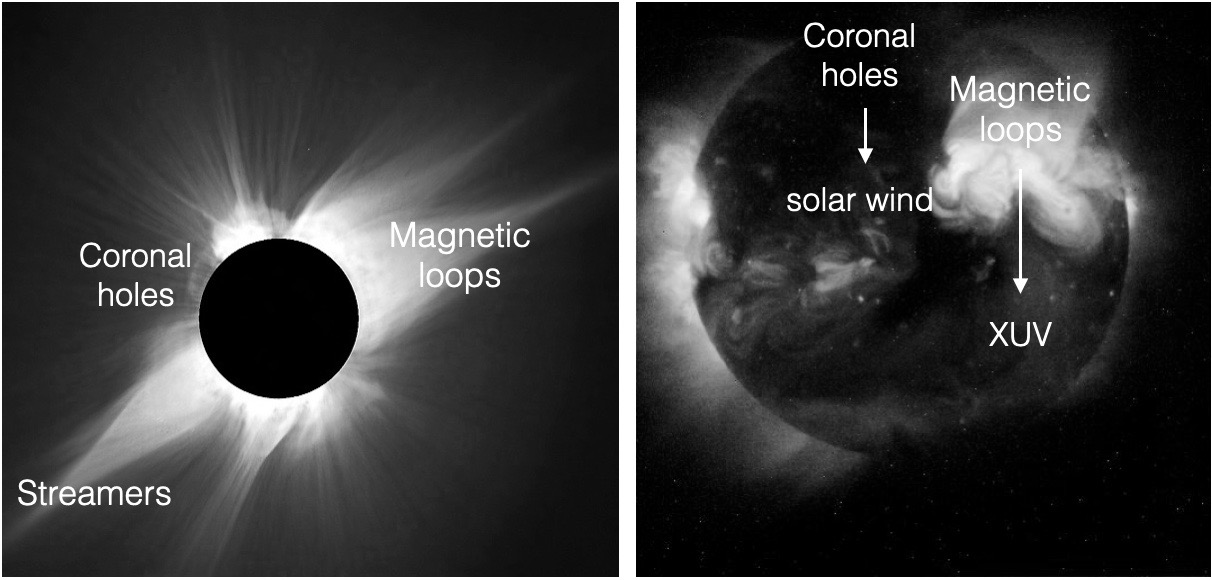} 
 \caption{
 Left: White light eclipse picture of the solar corona, showing how coronal magnetic field confinement in loops extends into outward
  steamers by the wind expansion, while open field coronal holes allow direct wind outflow.
 Right: Corresponding X-ray corona from emission by the MK plasma.
 Image credits: NASA. }
   \label{fig05}
\end{center}
\end{figure}

Measurements by interplanetary spacecraft give a typical solar wind speed of $v_\infty \approx 400$\,km/s,
 ranging up to $v_\infty \approx 700$\,km/s in high-speed streams thought to originate from open-field coronal holes;
these values thus do indeed straddle the solar escape speed, $v_{esc} \approx 618$\,km/s.

The observed mass flux implies to a quite small global mass loss rate, ${\dot M} \approx 10^{-14} M_\odot$/yr,
which is even less than the mass loss associated with the Sun's radiative luminosity, 
${\dot M}_{rad} \equiv L_\odot/c^2 \approx 5 \times 10^{-14} M_\odot$/yr.
Over the Sun's entire $\sim 10$\,Gyr lifetime, it will thus lose only $0.01\%$ of its mass via the solar wind.
The wind kinetic luminosity is ${\dot M} v_\infty^2 /2 \approx 10^{-7} L_\odot$, which turns out quite comparable the observed 
total coronal emission in X-ray and XUV, $L_{xuv}$.

As discussed in section 4, such coronal emission from cool stars with exoplanets can play an important role in heating planetary atmospheres 
and inducing their own wind outflows.

\subsection{Angular momentum loss and spindown: the Skumanich law}

While the mass loss from coronal winds from the Sun and other cool stars is negligible, the coronal magnetic field provides an extended
moment arm that can make the associated loss of angular momentum in the magnetized wind have a significant effect in causing an
evolutionary spindown in the star's rotation.  This is indeed thought to be the key cause of the relatively slow rotation period of the present-day Sun, $P_{rot} \approx 27$\,d.

Following pioneering analysis of \citet{WD67}, for a star of rotation frequency $\Omega$, the loss of stellar angular momentum ${\dot J}$
from a wind with mass loss rate ${\dot M}$ is given by
\beq
{\dot J} = (2/3) {\dot M} \Omega R_A^2
\,,
\label{eq:jdot}
\eeq
where the effective moment arm is set by the Alfv\'{e}n radius $R_A$, defined to be where the wind outflow
speed equals the Alfv\'{e}n speed, $V(R_A) \equiv  V_A \equiv B/\sqrt{4 \pi \rho}$.
For the present-day solar wind and magnetic field, one finds $R_A \approx 20 R_\odot$, which when applied in (\ref{eq:jdot})
gives a spindown time,
\beq
t_s \equiv \frac{J}{{\dot J}} \approx 10 \, {\rm Gyr}
\, .
\label{eq:tj}
\eeq
The fact that this comparable to the Sun's main sequence lifetime is consistent with the notion that Sun's present-day
slow rotation period of 27\,d is the result of wind spindown.

More quantitatively, observations of a large population of solar-type stars shows their rotation follow a simple relation,
known as the ``Skumanich law'' \citep{Skumanich19}.
From an initial period $P_o$, the period $P$ increases with the square root of the age $t$,
\beq
P(t) \approx P_o \left ( \frac{2t}{t_{so}} + 1 \right )^{1/2}
\, .
\label{eq:poft}
\eeq
If one assumes that the magnetic field scales with $B \sim \Omega \sim 1/P$, then
application of the \citet{WD67} analysis gives a simple scaling for the initial spindown time
with the initial field strength $B_o$,
\beq
t_{so} \approx 0.09 \, \frac{M V_w}{R^2 B_o^2} = 0.38 \frac{\rho R  V_{w} }{B_o^2} \approx 11.7 {\rm Gyr} \left ( \frac{B_o}{2 \,G} \right )^{-2}
\, ,
\label{eq:tswd}
\eeq
where $\rho$ and $R$ are the star's mean density and radius, and $V_{w}$ is its wind speed.
The last equality provides a numerical evaluation for solar parameters.
Remarkably, the dependence on wind mass loss rate scales out, through cancellation of the ${\dot M}$ in 
(\ref{eq:jdot}) with the inverse dependence in the Alfv\'{e}n radius, $R_A^2 \sim 1/{\dot M}$.

For example, taking the Sun's initial rotation period to be $P_o \approx 1$\,d,  achieving the present $P \approx 27$\,d for the Sun's current age $t \approx 4.6$\,Gyr
requires an initial field $B_o \approx 61$\,G and initial spindown time of just $t_s \approx 12.6$\,Myr.
This gives for the present global field $B_{now} = 61/27 = 2.3$\,G.

The upshot is that the young Sun was likely much more rapidly rotating, with stronger magnetic activity,  a stronger wind,
and greater coronal EUV emission.
These likely played a role in stripping the initially much denser atmosphere of Mars, implying that younger, more active stars might 
have similar effects on the atmospheres of the exoplanets, as we next discuss.

\begin{figure}[b]
\begin{center}
\includegraphics[width=2.35in]{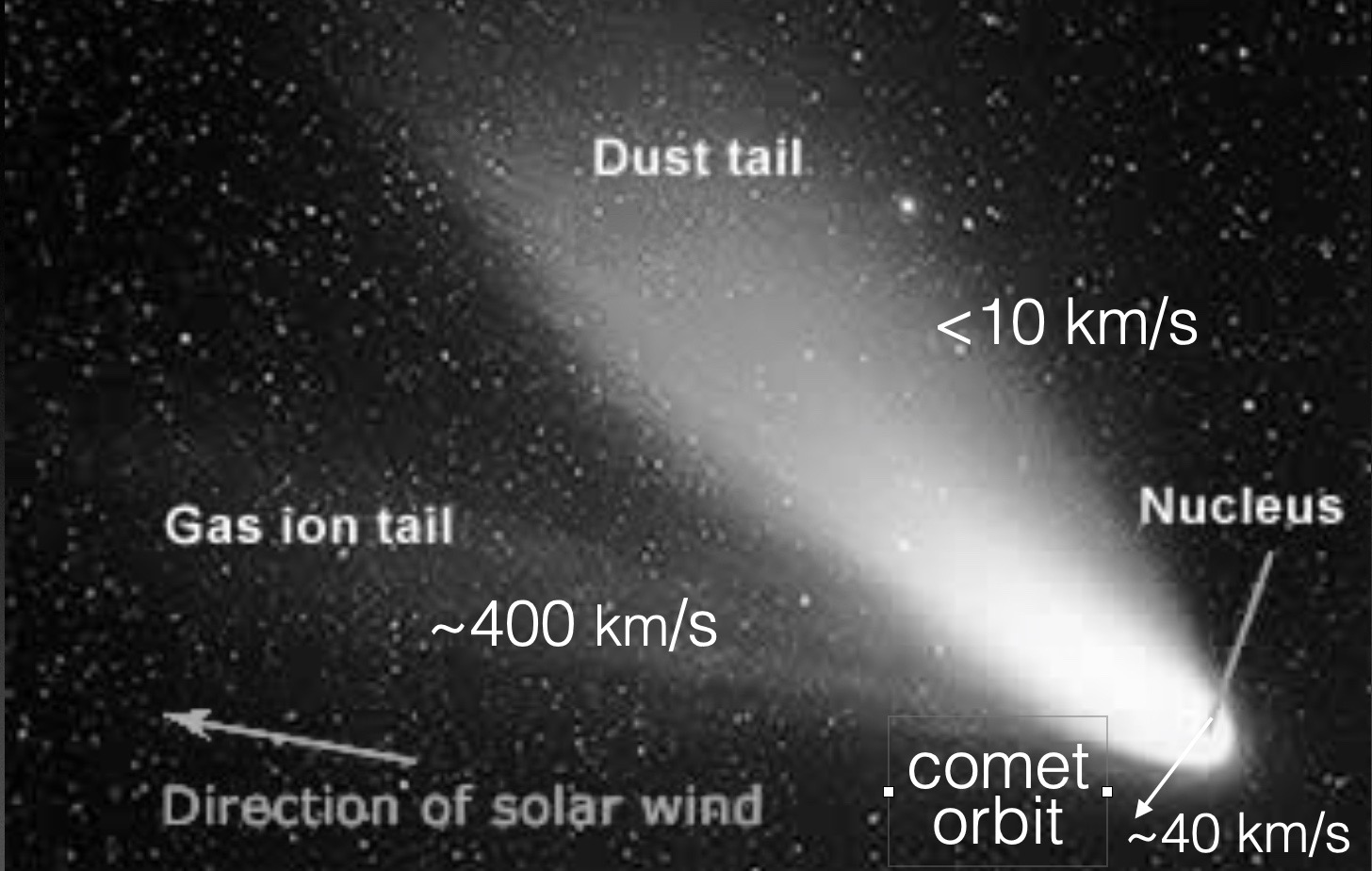} 
\includegraphics[width=2.65in]{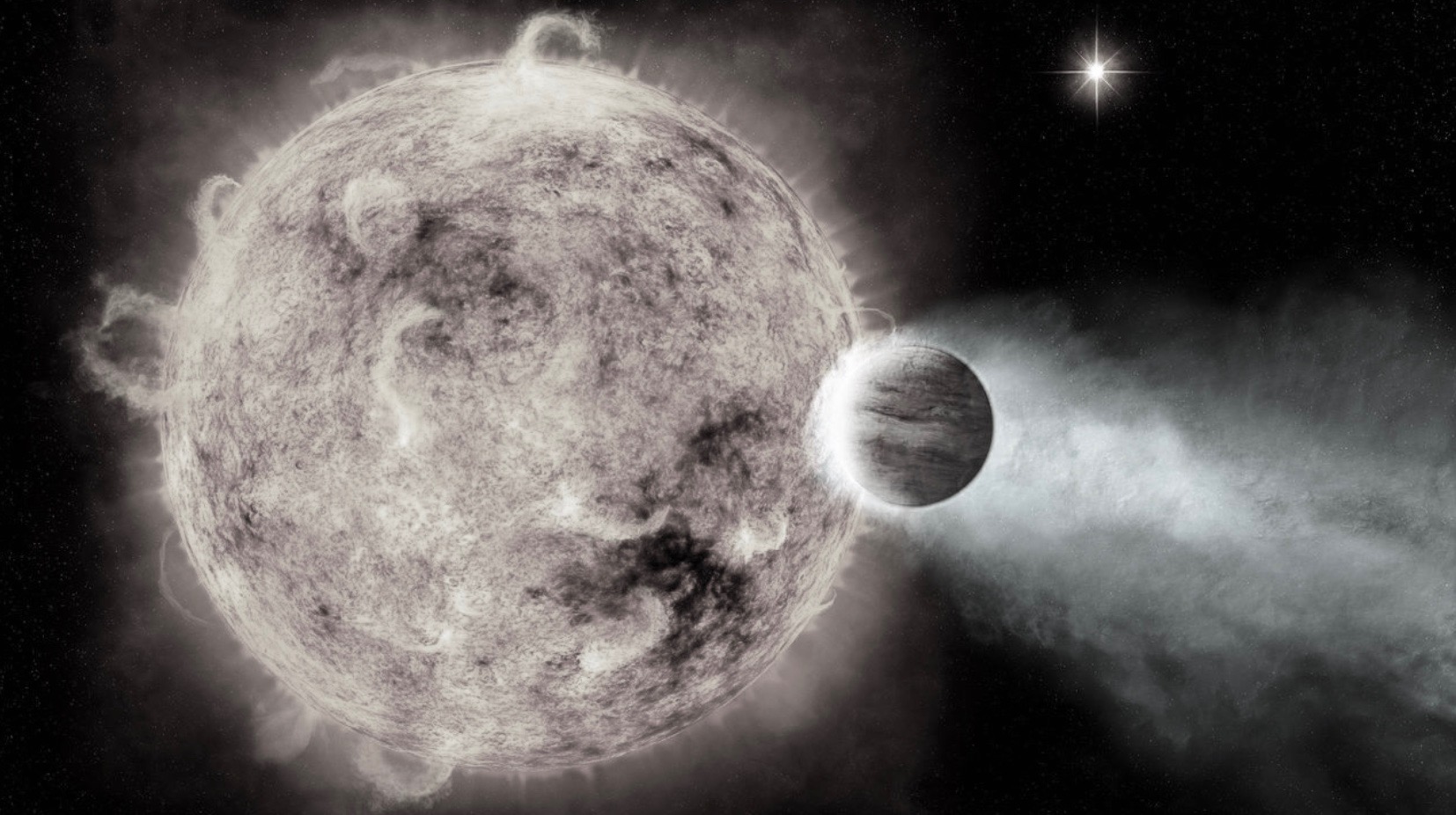} 
 \caption{
 Left: Illustration of dust and ion tails from comets. The former is driven  away slowly by solar radiation, thus trailing behind the comet's 
 orbital motion. The latter is driven by the much faster solar wind, and so is oriented nearly in an anti-solar direction.
 Right: illustration of mass loss from a hot-Jupiter exoplanet can be similarly driven away in a comet tail by interaction with the star's light and wind.
 Image credits: NASA.
  }
\label{fig06}
\end{center}
\end{figure}

\section{Planetary winds and mass loss}

The above discussion of solar and stellar wind mass loss from expansion of a hot corona provides a good basis for exploring planetary winds and the associated depletion of their atmospheres.
One key difference is that solar and stellar coronae form from inside-out heating from the star,
in contrast to the mainly external heating and ablation of a planetary atmosphere from the parent star and its wind.
As illustrated in figure \ref{fig06}, this can be expected to cause any planetary outflow to take the form of a cometary tail,
much as has been observed extensively for comets orbiting the Sun.

Unlike the more spherical, solar coronal expansion, which must become supersonic to match the outer boundary condition of a low
interstellar pressure, the cometary tail mass loss from planets could, in principal, be quite slow and subsonic, even representing the outer
layer loss of an otherwise nearly hydrostatically stratified atmosphere.
But even if the kinetic energy of the outflow thus remains small and negligible, there remains a key energy requirement for some source of heating to power any mass loss against the gravitational binding from the planet.

\subsection{Planetary escape temperatures}

In this context, it helpful to consider the general scaling and values of the {\em escape temperatures} of planetary bodies as a function of their mass $M$ and radius $R$, and the molecular weight $\mu$ (in units of proton mass $m_p$) of their atmospheres,
\beq
T_{esc} \equiv  \frac{GM/R}{k/\mu m_p}  = 131 \, {\rm kK} \, \frac{M/M_J}{R/R_J} 
= 131 \, {\rm kK} \, \frac{\rho}{\rho_J}  \, \left ( \frac{R}{R_J} \right )^2
\, .
\label{eq:TescJ}
\eeq
Here the last two equalities give scalings relative to Jupiter, assuming $\mu = 0.6$ that applies for a fully ionized mix
of H and He at solar abundances.
For gas giants, the mean density $\rho$ is (like the Sun) typically of order unity CGS (e.g., $\rho_J \approx 1.3$\,g/cm$^3$), and so the last equality shows that the escape temperature  mainly scales quadratically with the planetary radius.

For smaller, rocky, terrestrial planets like the Earth, the density is a bit higher ($\rho_E \approx 5.5$\,g/cm$^3 \approx 4.1 \rho_J$,)  but the radius is much smaller ($R_E \approx R_J/10$),
giving the Earth an escape temperature of about $4500$\,K.

By comparison, if we approximate planetary absorption and emission as a simple blackbody,
then for a planet at a distance $d$ from a star with effective temperature $T_\ast$ and radius $R_\ast$,
the equilibrium temperature is given by
\beq
T_{eq} = 290 \, {\rm K} \, \frac{T_\ast}{T_\odot} \sqrt{ \frac{R_\ast/R_\odot}{d/{\rm au}} }
= 4250 \, {\rm K} \, \frac{T_\ast}{T_\odot} \sqrt{ \frac{R_\ast}{d} }
\, .
\label{eq:Teq}
\eeq
The first equality is cast in terms of interplanetary distances like that of the Earth at 1\,au;
it actually matches quite well the case of Earth's actual mean temperature, but this is due to a rather
fortuitous cancelation between the effect of the high albedo of clouds in reducing the absorption of solar radiation,
and the greenhouse effect in trapping Earth's own cooling radiation.

The latter equality frames this in terms of very close-in planets, like the so-called ``hot Jupiters".
A key point here is that even for close-in planets, the direct  heating from the star's thermal radiation
is almost never sufficient to bring a planet anywhere near the typical escape temperature associated
with the thermal wind expansion from the planet.

Instead, it is the much harder, XUV radiation from stellar coronae that can lead such escape-level heating, as we next discuss.

\subsection{Planetary mass loss from heating by XUV coronal radiation} 

For photons ($\gamma$) of a given wavelength $\lambda$ and thus energy $E = h \nu = h c/\lambda$,
we can cast this energy in terms of an associated temperature,
\beq
T_\gamma \equiv \frac{h \nu}{k} = 11.6 \, {\rm kK} \, \frac{h\nu}{eV} = 14 \, {\rm kK} \, \frac{1 \mu m}{\lambda}
\, .
\label{eq:Tgamma}
\eeq
For photons above the Hydrogen ionization limit with $h \nu > 13.6$\,eV and $\lambda < 91.2$\,nm, the associated
escape temperature is $T_{esc} > 156$\,kK.
Coronal emission at EUV energies can thus readily ionize H with a sufficient excess energy to heat the gas to
temperature well above typical escape temperature.

\begin{figure}[t]
\begin{center}
\includegraphics[width=2.35in]{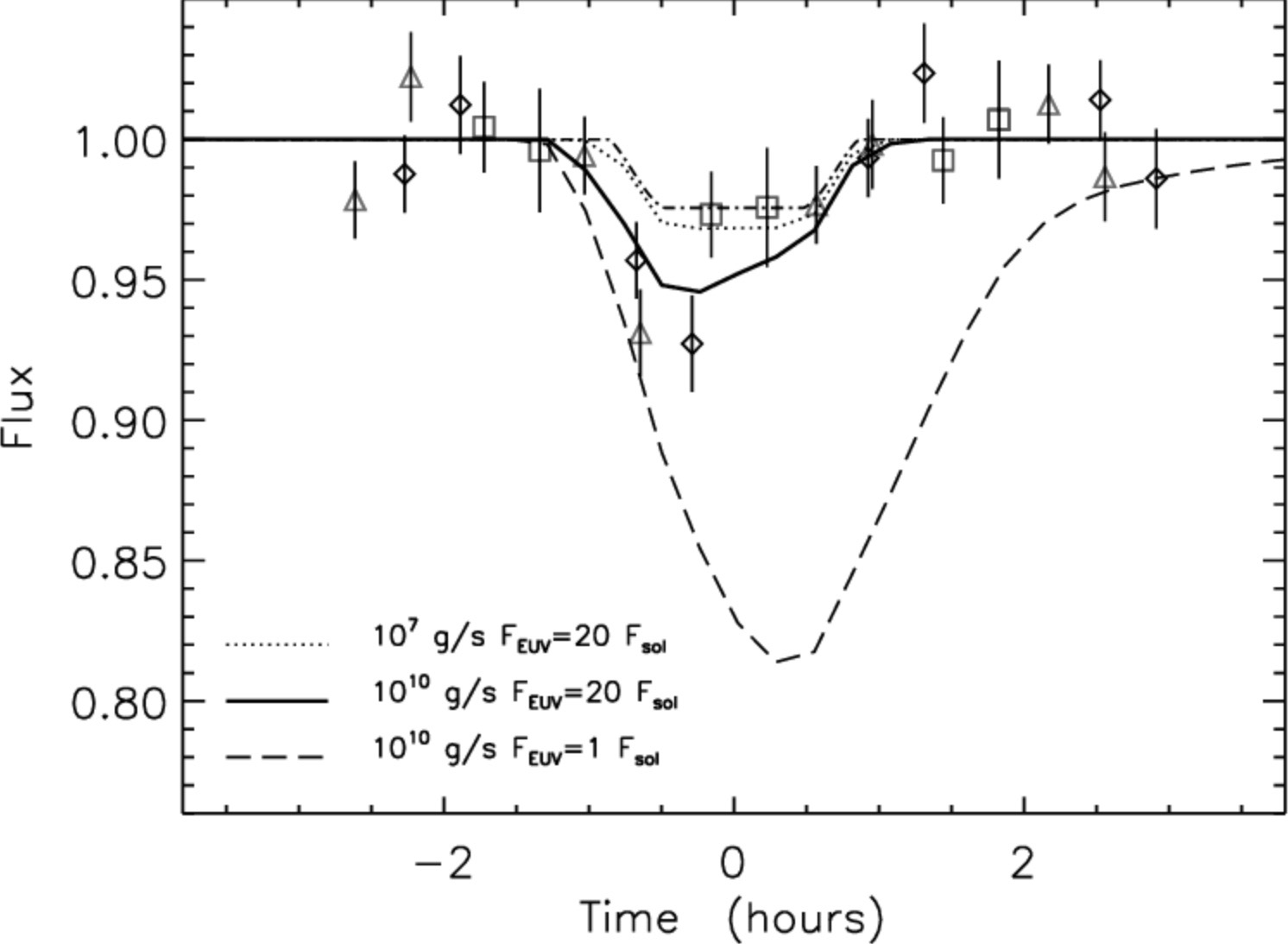} 
\includegraphics[width=2.65in]{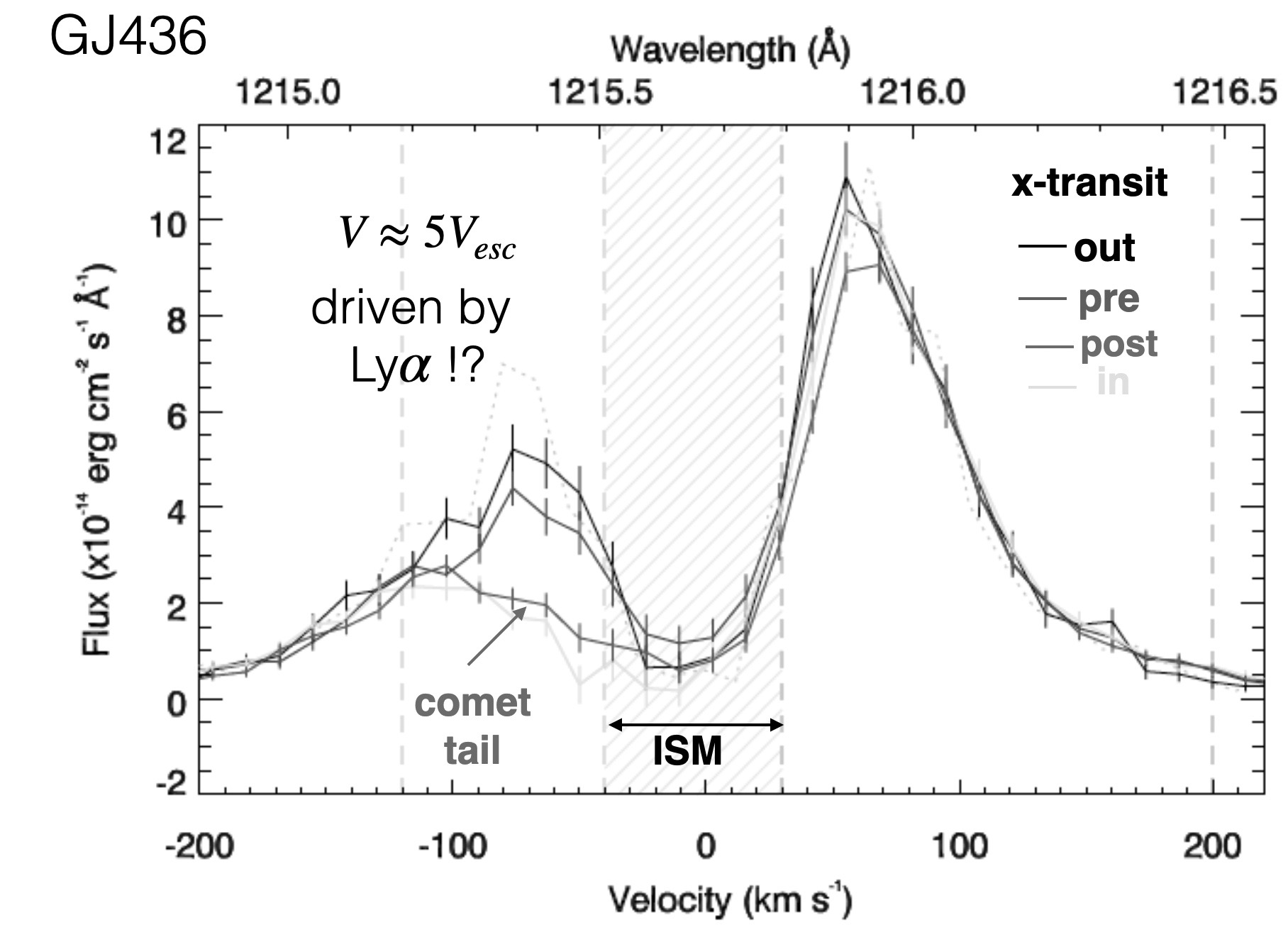} 
\vspace{-0.1in} 
 \caption{
 Left: Ly$\alpha$ transit light for hot  Jupiter  HD1897333b, comparing data points with values models with various assumed mass loss rates and EUV fluxes.  Figure from \citet{LDE10}.
 Right: Ly$\alpha$ transit spectrum for warm Neptune GJ436, overplotted for 4 phases in which the planet was pre, post, in and out of transit.
 Note, in particular, that the post-transit spectrum shows evidence for a comet tail, with a speed that likely reflects either radiative acceleration by stellar Ly$\alpha$ or entrainment in the outflowing stellar wind, rather than acceleration by thermal expansion from the planet.
 Adapted from Ehrenreich et al. (2015).
  }
\label{fig07}
\end{center}
\end{figure}
In analogy with the scaling in eqn. (\ref{eq:mdotq}) that was derived for mass loss from stellar coronae, the associated 
planetary mass loss  depends on the net  XUV heating of its outer atmosphere.
For a planet of mass $M_p$ and radius $R_p$ at a distance $d$ from a star with coronal XUV luminosity that is a fraction
of  $f_{xuv}  \equiv L_{xuv}/L_\ast$ of the stellar luminosity,
the energy balance to drive a planetary mass loss ${\dot M}$ is
\beq
{\dot M} \frac{G M_p}{R_p} \approx \epsilon \, f_{xuv} \, \frac{L_\ast}{4 \pi d^2} \, \pi R_p^2
\, ,
\label{eq:escen}
\eeq
where $\epsilon$ is an order-unity efficiency factor for this XUV heating.
Solving for the mass loss shows that it scales with the ratio of XUV flux to the planet's mean density,
\beq
{\dot M} = 1.5 \epsilon f_{xuv} \, \frac{L_\ast/4\pi d^2}{G \rho_p}
\, .
\label{eq:mdotp}
\eeq

For example, assuming $\epsilon \approx 1$ and a solar value $f_{xuv} \approx 10^{-7}$ for XUV fraction, 
we find  a hot Jupiter at distance $d=2R_\odot$ from a solar-type
star should have mass loss rate ${\dot M} \approx 10^{10}$\,g/s.
The associated mass loss time is $t_{atm} \equiv  M/(dM/dt) \approx 4 \times 10^{12}$\,yr.
A central conclusion is thus that even for such an extreme case, the total mass loss from such gas giants is small even over
the evolutionary lifetime of the system.

In contrast, for terrestrial planets like Earth, the atmosphere is a tiny fraction $f_{atm} \approx 10^{-6}$ of the planet's mass.
Even at the greater distance $d=1$\,au, the associated  time for Earth to loss a Hydrogen atmosphere is
$t_{atm} \approx 670$\,Myr, and even shorter if one accounts for the likely stronger coronal emission from the early Sun.

By contrast, the present-day atmosphere of Nitrogen and Oxygen has been mostly retained, because of its higher molecular weight,
although there is evidence of a weak ``polar wind" mass loss from polar regions of open magnetic fiield.

As illustrated in figure \ref{fig07}, mass loss from exoplanets can be observationally diagnosed through transit light curves and spectra,
often centered as here on the UV  Lyman-$\alpha$ transition of Hydrogen at $\lambda \approx 91.2$\,nm.
The results here are for two hot giants, with the light curve in the left panel providing constraints on the planets mass loss and star EUV flux.
The overplot of the Ly$\alpha$ spectral line at four phases of the transit provides information on the outflow speed and its orientation in a comet tail extending away from the star.

\section{Magnetospheres of stars and planets}

\begin{figure}
\begin{center}
\includegraphics[width=5.35in]{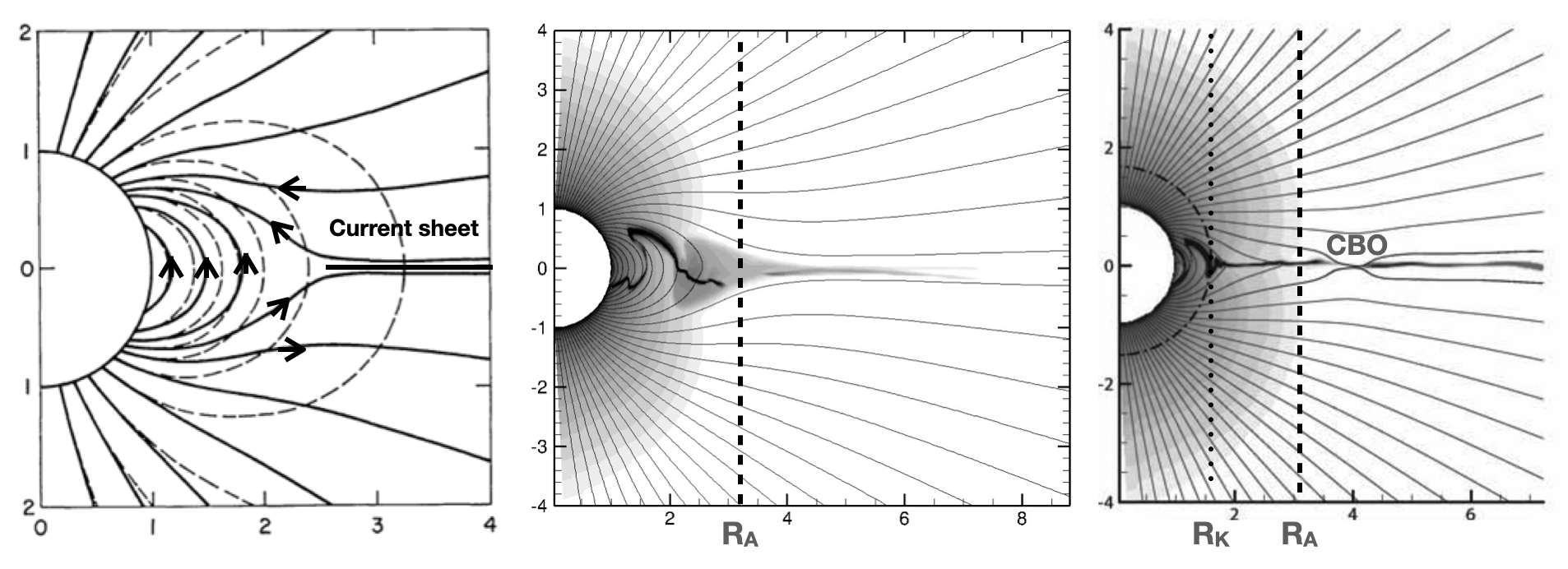} 
\caption{
 Left:  Iterative solution from \citet{PK71} for how an initial dipole field (dashed lines) in the solar corona
 is by altered by gas pressure and solar wind expansion into a stretched and eventually open form, with lines
 of opposite polarity (denoted by arrows) separated by a current sheet.
 Middle:  Snapshot of MHD simulation of analogous distortion of dipole field from a magnetic hot star, 
 this time by its strong, radiatively driven wind, with the Alfv\'{e}n radius $R_A $ marking the transition
 from inner closed loops to radially stretched field lines in the outer wind. 
 Right: Analogous MHD snapshot for a hot star with rapid rotation, giving a Kepler co-rotation radius $R_K \ll R_A$,
 forming a wind-fed CM in the region between, punctuated by episodic CBO events, with associated magnetic 
 reconnection.
   }
\label{fig08}
\end{center}
\end{figure}

Mass loss from both stars and planets can be strongly affected by magnetic fields.
Figure \ref{fig05} illustrates vividly the key role of fields in structuring the solar coronal 
X-ray emission and the solar wind outflow.
The white-light coronal image on the left shows how gas trapped by closed magnetic loops 
near the Sun gives way to a radially pointed streamer structures.
As illustrated in left panel of figure \ref{fig08}, the basic mechanism was first demonstrated in a pioneering analysis by 
\citet{PK71},
 who examined how an initially closed magnetic dipole in the solar corona is forced open 
by gas-pressure expansion. 
The opening of opposite polarity field lines induces a current-sheet in the outer, expanding wind,
where they are a central feature of {\em in situ} spacecraft measurements.
Analogous magnetic effects likely occur other cool stars with hot coronal winds,
 but the complexity of these dynamo-generated fields leads to significant cancellation in disk-integrated light, 
making them difficult to diagnose remotely through spectro-polaritmetry.

\subsection{Massive-star magnetospheres}

In contrast,  although massive, luminous, hot stars lack the hydrogen recombination
convection zone that induces the magnetic dynamo cycle of cooler,
solar-type stars, modern spectropolarimetry has nonetheless
revealed that about 10\% of O, B and A-type stars harbor
large-scale, organized (often predominantly dipolar) magnetic fields
ranging in dipolar strength from a few hundred to tens of thousand Gauss.
These fields, which are likely fossils of an earlier epoch, channel and trap the strong, radiatively driven  winds of such stars,
feeding a circumstellar magnetosphere.

The inside-out building of these wind-fed magnetospheres is in
some way complementary to the outside-in nature of the
planetary magnetospheres impacted the star's wind. But
there are also some interesting similarities in the role of the
characteristic magnetospheric and corotation radii.

\subsubsection{ Alfv\'en radius and Kepler co-rotation radius}

MHD  simulation studies
\citep[e.g.,][]{uDO02, Uddoula08}
 show that the overall net effect of a large-scale,
dipole magnetic field in diverting such a hot-star wind can be
well characterized by a single \textit{wind magnetic confinement
parameter} and its associated \textit{Alfv\'en radius},
\begin{equation}
\eta_{\ast} \equiv \frac {B_{eq}^2 \, \rs^2} {\mdot \, \vinf} ~~ ;
~~ \frac{\ra}{\rs} \approx 0.3 + \left ( \etas + 0.25
\right)^{1/4} \, ,
\label{eq:esdef}
\end{equation}
where $\Beq = \bp/2$ is the field strength at the magnetic
equatorial surface radius $\rs$, and $\mdot$ and $\vinf$ are the
fiducial mass-loss rate and terminal speed that the star
\textit{would have} in the \textit{absence} of any magnetic field.
 This confinement parameter sets the scaling for
the ratio of the magnetic to wind kinetic energy density. For a
dipole field, the $r^{-6}$ radial decline of magnetic energy
density is much steeper than the $r^{-2}$ decline of the wind's
mass and energy density; this means the wind always dominates
beyond the Alfv\'en radius, which scales as $\ra \sim \etas^{1/4}$
in the limit $\etas \gg 1$ of strong confinement.

\begin{figure}[t!]
    \begin{center}
\includegraphics[width=120mm]{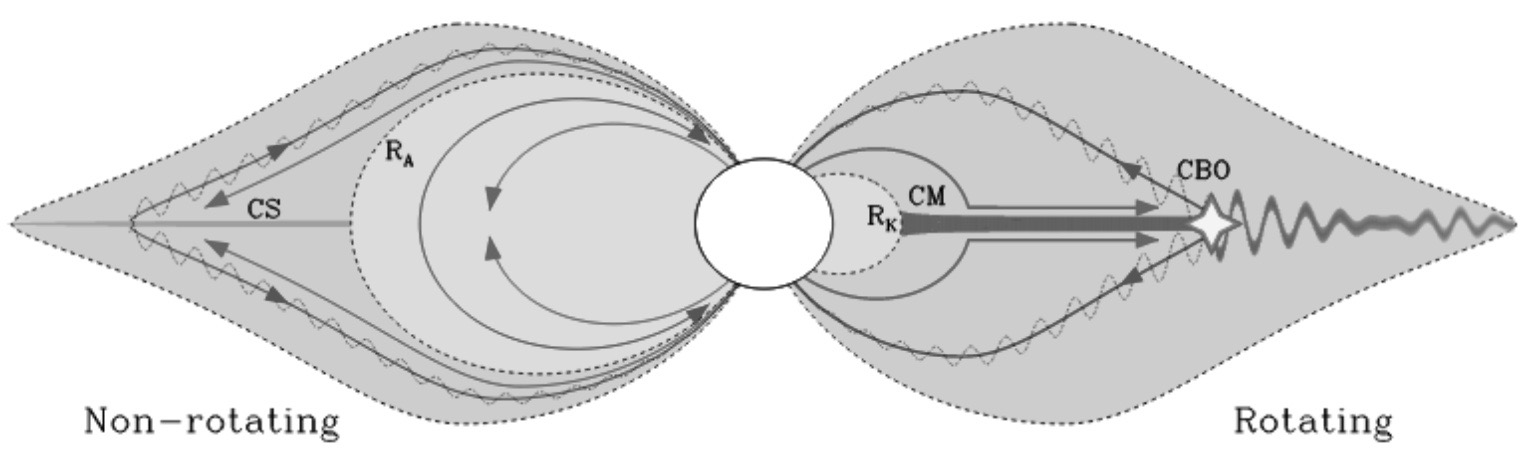}
\vspace{-0.1in} 
 \caption{
 Schematic of stellar magnetospheres, contrasting those from non-rotating (or slowly rotating) stars (left)
 from rapidly rotating stars (right).
 On the left $R_A$ denotes the Alfv\'{e}n radius, separating the region of closed magnetic loops from
 the magnetized outflowing wind, which stretches the field  into open lines of opposite polarity, separated by a current sheet (CS).
 On the right $R_K$ denotes the Kepler co-rotation radius, above which the centrifugal force exceeds gravity,
 trapping wind material in a centrifugal magnetosphere (CM), until the mass build-up overwhelms the magnetic 
 field confinement, leading to centrifugal breakout (CBO) events.
 }
 \label{fig09} 
\end{center}
\end{figure}

For a model with $\etas = 100$, the middle panel of figure \ref{fig08} shows
 magnetic loops extending above
$\ra \approx 3.2 \rs$ are drawn open by the wind, while those with an apex below
$\ra$ remain closed,  trapping wind upflow from
opposite footpoints of closed magnetic loops.
Once this material cools back to
near the stellar effective temperature, it falls back onto the star over a dynamical timescale.

The dynamical effects of rotation can be analogously parameterized
\citep{Uddoula08}
 in terms of the {\em orbital rotation
fraction}, and its associated {\em Kepler corotation radius},
\begin{equation}
W \equiv \frac{V_{\rm rot}}{{V_{\rm orb}}} =  \frac{V_{\rm
rot}}{\sqrt{G\Mstar/\Rstar}} ~~ ; ~~ \Rkep = W^{-2/3}\,\Rstar
\label{eq:rkep}
\end{equation}
which depends on the ratio of the star's equatorial rotation speed
to the speed to reach orbit near the equatorial surface radius
$\Rstar$. Insofar as the field within the Alfv\'en radius is
strong enough to maintain {\em rigid-body rotation}, the Kepler
corotation radius $\Rkep$ identifies where the centrifugal force
for rigid-body rotation exactly balances the gravity in the
equatorial plane. 
Figure \ref{fig09} contrasts the roles of magnetic confinement
and centrifugal support in stellar magnetospheres.

\begin{figure*}
\begin{center}
    \includegraphics[width=130mm]{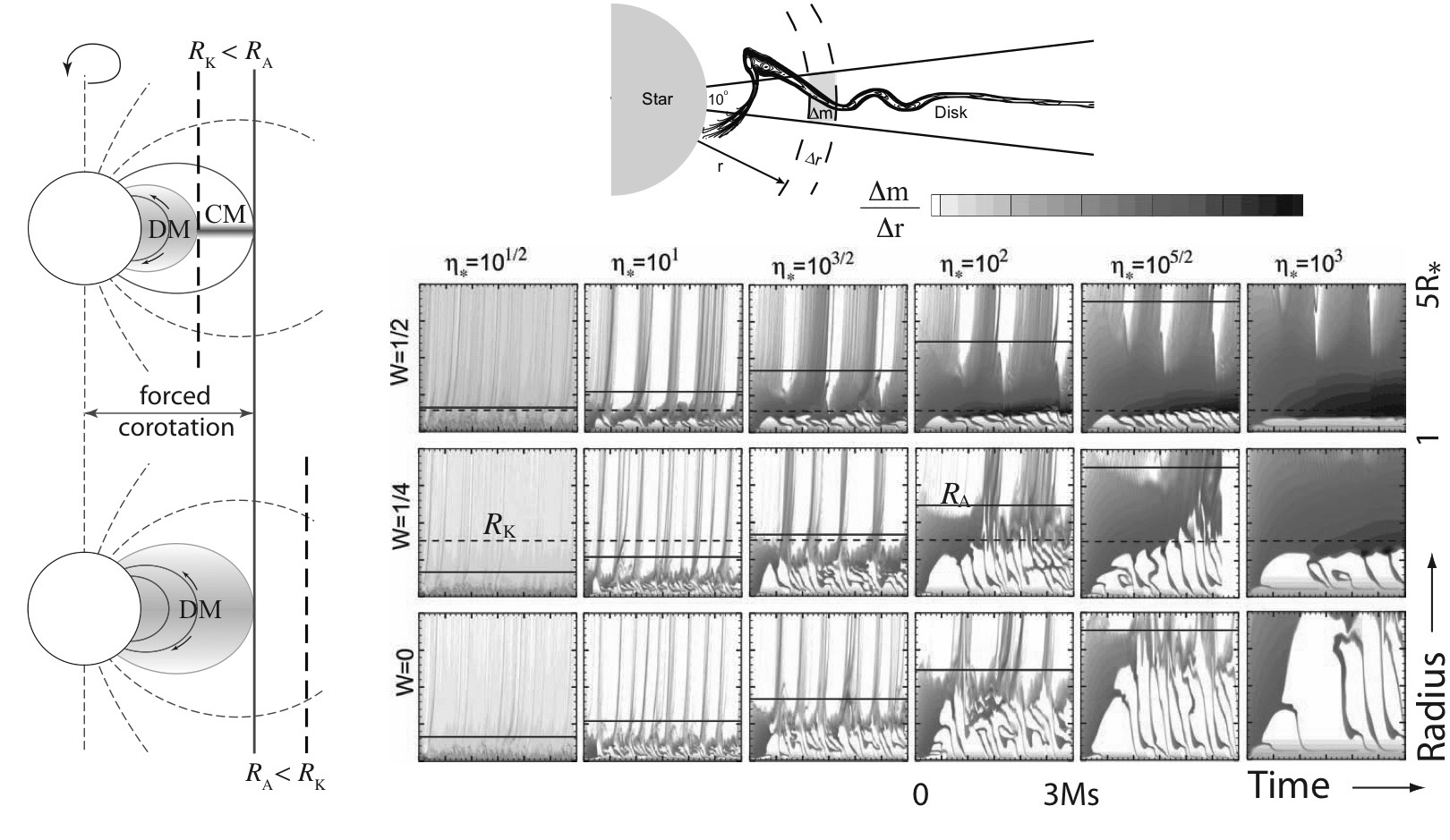}
  \end{center}
  \vspace{-0.1in} 
\caption{ \small{ {\em Left:} Sketch of the regimes for a
dynamical vs.\ centrifugal magnetosphere (DM vs. CM). The lower
panel illustrates the case of a slowly rotating star with Kepler
radius beyond the Alfv\'{e}n radius ($\rk > \ra$); the lack of
centrifugal support means that trapped material falls back to the
star on a dynamical timescale, forming a DM, with shading
illustrating the time-averaged distribution of density. The upper
panel is for more rapid rotation with $\rk < \ra$, leading then to
a region between these radii where a net outward centrifugal force
against gravity is balanced by the magnetic tension of closed
loops; this allows material to build up to  the much higher
density of CM. {\em Right, Upper:}
 Contour plot for density at arbitrary snapshot of an isothermal 2D MHD simulation with magnetic confinement parameter $\estar=100$ and critical rotation factor $W=1/2$.
 The overlay illustrates the definition of radial mass distribution, $\Delta m/\Delta r$, within $10^\circ$ of the equator.
{\em Right, Lower:} Density plots for  log of $\Delta m/\Delta r$,
plotted versus radius (1-5 \Rstar) and time (0-3~Msec), for a
mosaic of 2D MHD models with a wide range of magnetic confinement
parameters \estar, and 3 orbital rotation fractions $W$. The
horizontal solid lines indicate the Alfv\'en radius \Ralf\ (solid)
and the horizontal dashed lines show  Kepler radius \Rkep\
(dashed). } 
\label{fig10} }
\end{figure*}

\subsubsection{Dynamical vs.\ Centrifugal Magnetosphere}

If $\Ralf < \Rkep$, then material trapped in
closed loops will again eventually fall back to the surface on a dynamical timescale,
thus forming what's known as a {\em dynamical magnetosphere} (DM). 

But if $\Ralf >
\Rkep$, then wind material located between \Rkep\ and \Ralf\ can
remain in static equilibrium,
 forming a {\em centrifugal magnetosphere} (CM) that is supported against gravity by the magnetically enforced co-rotation.
As illustrated in the upper left schematic in figure
\ref{fig10}, the much longer confinement time allows material
in this CM region to build up to a much higher density than in a
DM region.

Eventually the centrifugal force of this material overwhelms the confining effect of magnetic tension,
leading to {\em centrifugal breakout} (CBO) events, as illustrated by the right panel of figure \ref{fig08}
for the case $\estar =100$ and $W=0.5$. (See also the right side of figure \ref{fig09}.)

For general 2D MHD simulations in the axisymmetric case of a
rotation-axis aligned dipole, the mosaic of greyscale plots in figure
\ref{fig10} shows the time vs. height variation of the
equatorial mass distribution $\Delta m/ \Delta r$ for various
combinations of rotation fraction $W$ and wind confinement
$\estar$ that respectively increase upward and to the right. This
illustrates vividly the DM infall for material trapped below
$\Rkep$ and $\Ralf$, vs.\ the dense accumulation of a CM from
confined material near and above $\Rkep$, but below $\Ralf$.
For such CM stars, note also the episodic CBO events.

\begin{figure}[t!]
    \includegraphics[width=120mm]{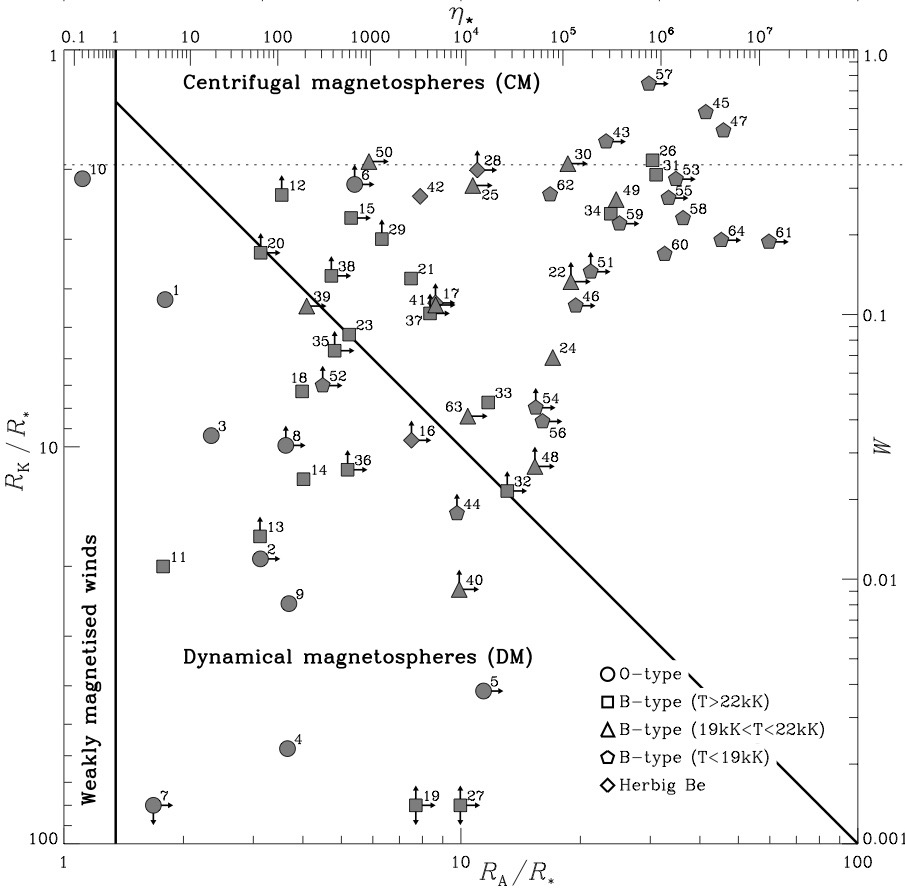}
\vspace{-0.1in} 
\caption{Classification of 64 observationally confirmed
magnetic massive stars in terms of  magnetic confinement vs.\
rotation fraction, characterized here by a log-log plot of Kepler
radius $\Rkep$ increasing downward vs.\ Alfv\'{e}n radius $\Ralf$
increasing to the right. The labeled ID numbers are sorted in
order of decreasing effective temperature $\teff$, with stellar identities given in
Table 1 of \citet{Petit13}.
Stars to the left of the vertical solid line have only weakly magnetized winds (with $\etas<1$),
Star below and left of the diagonal solid line have
dynamical  magnetospheres (DM) with $\Ralf<\Rkep$,
while those above and right of this line have centrifugal
magnetospheres (CM) with $\Ralf>\Rkep$.
}
 \label{fig11}
\end{figure}

\subsubsection{Comparison with Observations of Confirmed Magnetic Hot-stars}

For observationally confirmed magnetic hot-stars with $\teff
\gtrsim 16$\,kK, figure
\ref{fig11} plots positions in a log-log plane of $\rk$ vs. $\ra$ 
\citep{Petit13}.
The vertical solid line representing $\estar =1$
separates the domain of non-magnetized or weakly magnetized winds
to left,  from the domain of stellar magnetospheres to the right.
The diagonal line representing $\rk = \ra$ divides the domain
of centrifugal magnetospheres (CM) to the upper right from that
for dynamical  magnetospheres (DM) to the lower left. 

\begin{figure}[t!]
\vspace{-0.0in}
\begin{center}
   \includegraphics[width=120mm]{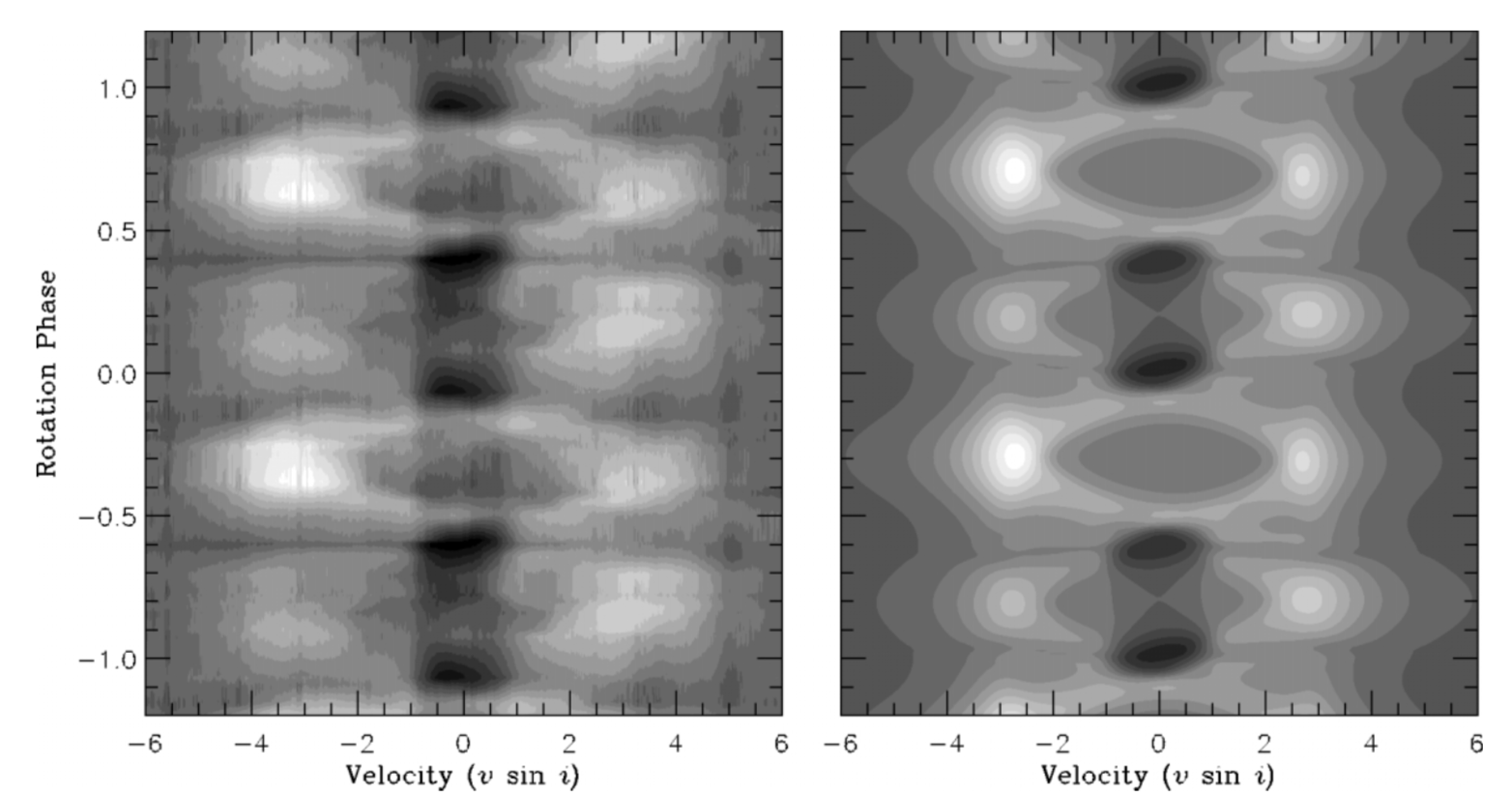}
\end{center}
\vspace{-0.1in} 
\caption{
Greyscale renditions of dynamic spectrum for variation of 
doppler-shifted emission and absorption with rotation phase.
The right panel shows results from an RRM model assuming rotation and
parameters quoted in the text for the strong CM star $\sigma~$Ori~E.
This is in good overall agreement with the left panel showing observed 
variations in H$\alpha$ from this star.
Adopted from \citet{TOG05}.
  }
  \label{fig12}
\end{figure}

For strong CM stars with $\rk \gg \ra$ one can apply a semi-analytic, rigidly rotating magnetosphere
(RRM) model developed by \citet{TO05}, which assumes the wind-fed mass accumulates along surfaces 
of minimum combined gravitational-centrifugal potential.
The strong confinement leads to a high enough density to make
the CM have a net emission in the Hydrogen Balmer line, H$\alpha$.
In the common case that the magnetic dipole axis has a non-zero titl
angle $\beta$ with the rotation axis, the highest density and strongest
emission comes from circumstellar clouds near the common rotational
and magnetic equator.
As rotation brings these clouds from in front of the star to off the limb,
the H$\alpha$ line profile shows corresponding variations from absorption
near line center to emission in the blue and red wings.
As shown in figure \ref{fig12}, the overall line-profile variation 
can be characterized by a {\em dynamic spectrum} that renders on 
a grayscale the  rotational phase variations of doppler-shifted emission and absorption.

For example, the strong CM star $\sigma~$Ori~E has a nearly dipole
field of  polar strength $B_p > 10^4$\, G and strong tilt $\beta > 60^o$,
and rotation fraction $W \approx 0.25- 0.5$.
The corresponding confinement parameter $\estar > 10^6$ implies
an Alfv\'{e}n radius  $\ra \approx  30 \rs $ well beyond the Kepler radius 
$\rk \approx 2 \rs $.
Figure \ref{fig12} compares dynamical spectra from a RRM model
(left) with that obtained from actual H$\alpha$ observations of $\sigma~$Ori E
(right).
The good overall agreement provides strong evidence for the overall CM
paradigm, as well as the basic RRM analysis method.

\begin{figure}[t!]
    \begin{center}
\includegraphics[width=56mm]{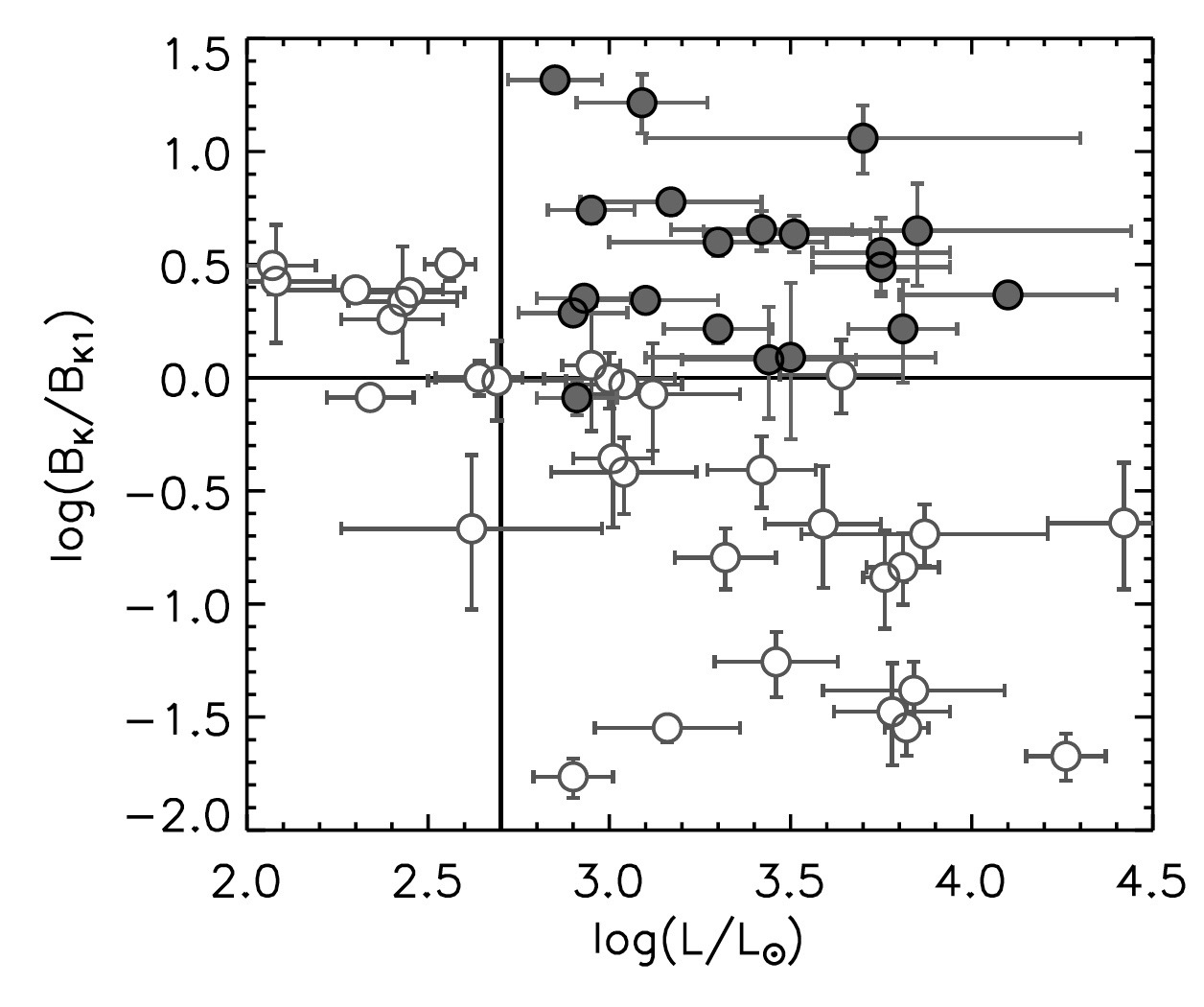}
\includegraphics[width=64mm]{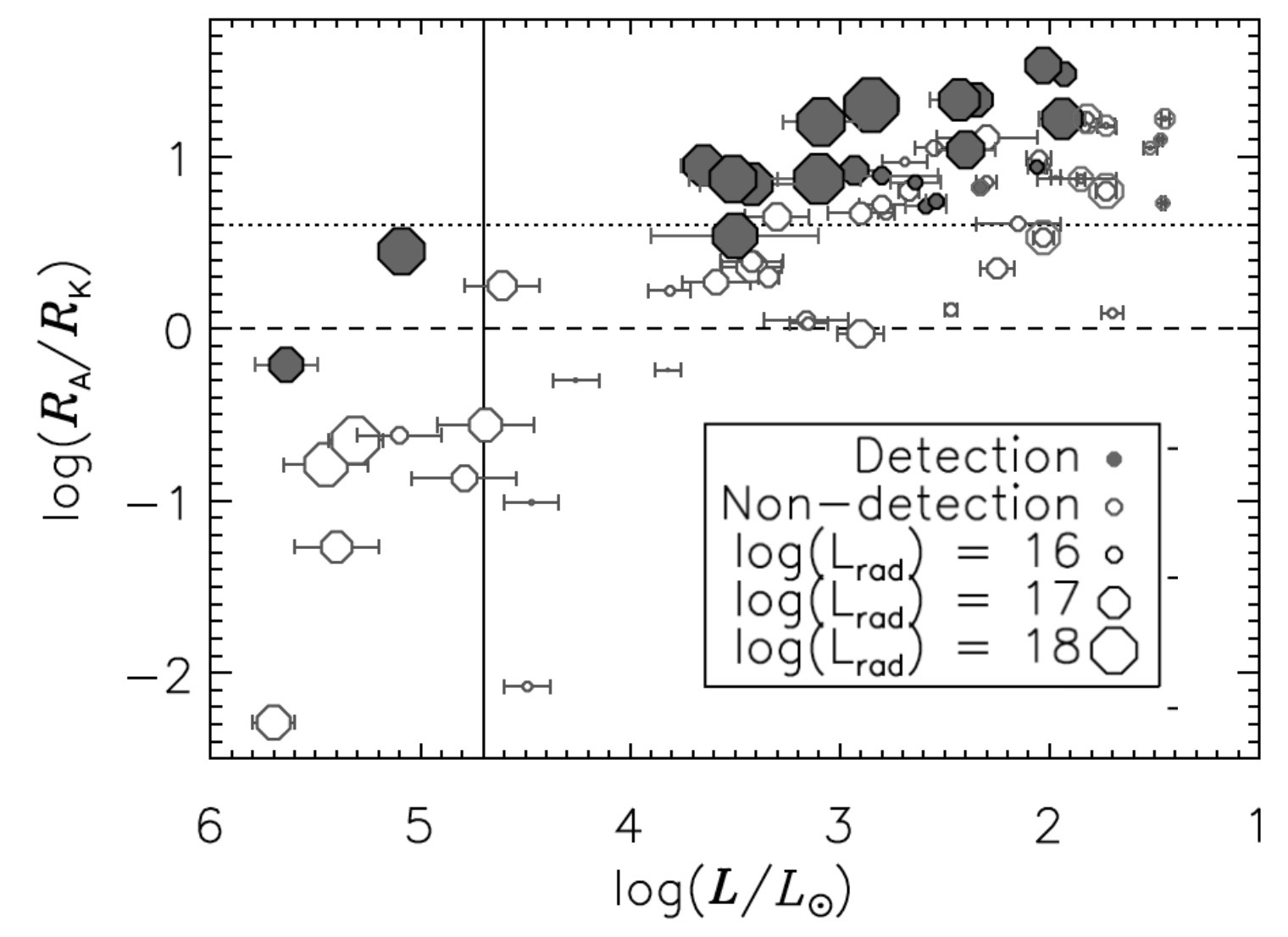}
\vspace{-0.1in} 
\caption{Left: Observed magnetic B-stars plotted in the $\log B_K/B_{K1}$ vs.\ $\log L/L_\odot$ plane, with stars showing H$\alpha$ in filled circles, and those without in open circles 
\citep{Shultz20}.
Here $B_K$ is the observationally inferred field strength at the Kepler co-rotation radius $\rk$, while is $B_{K1}$ is the strength the CBO model predicts is needed to make the CM have unit optical thickness in H$\alpha$ at $\rk$.
Above a threshold luminosity, the onset of H$\alpha$ emission occurs right at this CBO value, independent of luminosity;
this indicates the associated wind feeding of the CM overwhelms any leakage, to build the density to its breakout values.
Right:  Occurrence of radio emission for magnetic B stars in a log-log plane of the ratio $\ra/\rk$ vs. luminosity (now increasing from right to left).
Filled circles with detected radio lie the upper diagram, indicating rapid rotation again plays a key role in radio emission;
 the lack of clear dependence on luminosity again indicates the wind feeding rate is relatively unimportant,
 as expected for emission tied to CBO events and their associated magnetic reconnection.}
 \label{fig13} 
\end{center}
\end{figure}

\begin{figure}[t!]
    \begin{center}
\includegraphics[width=80mm]{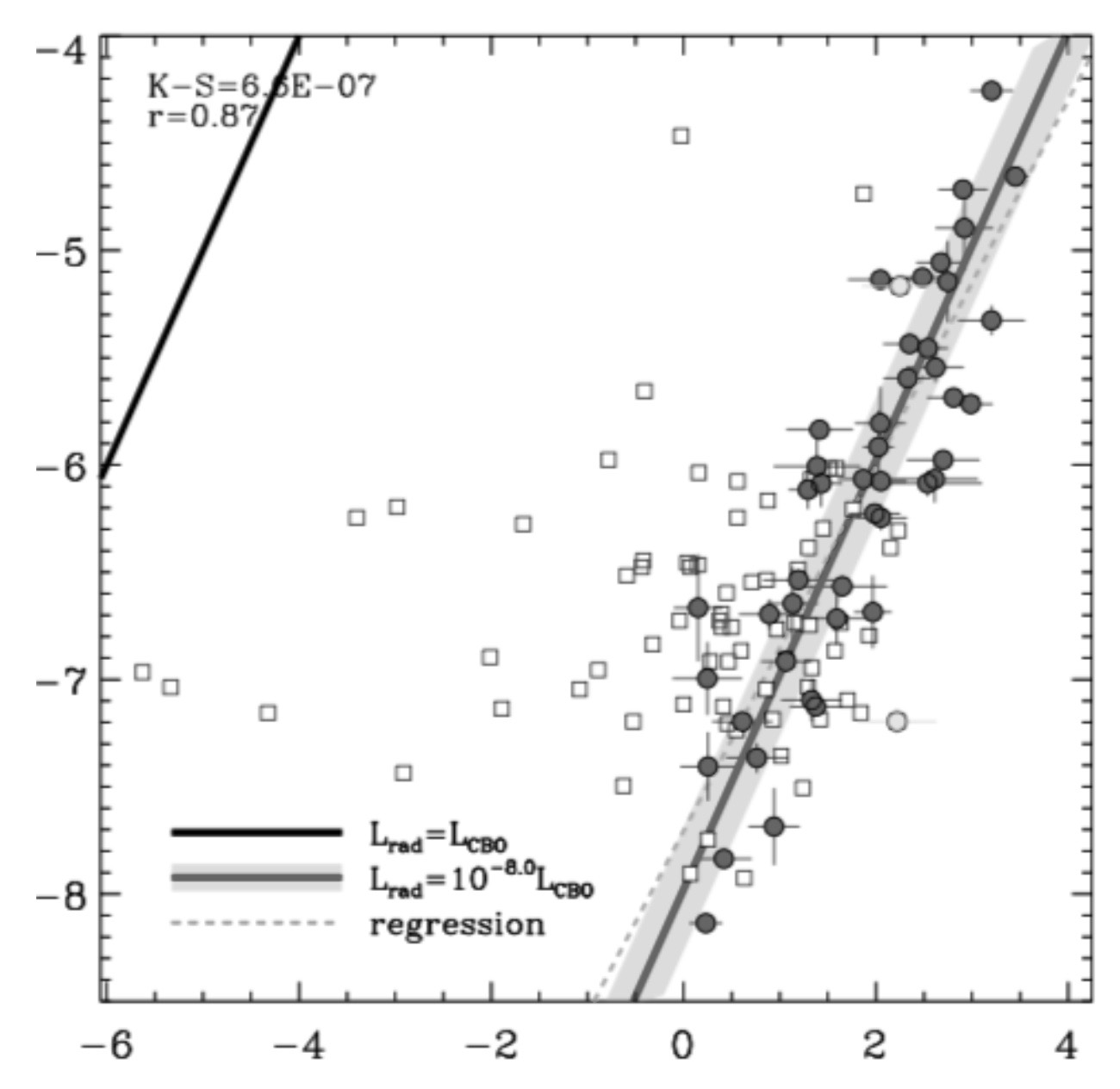}
\includegraphics[width=45mm]{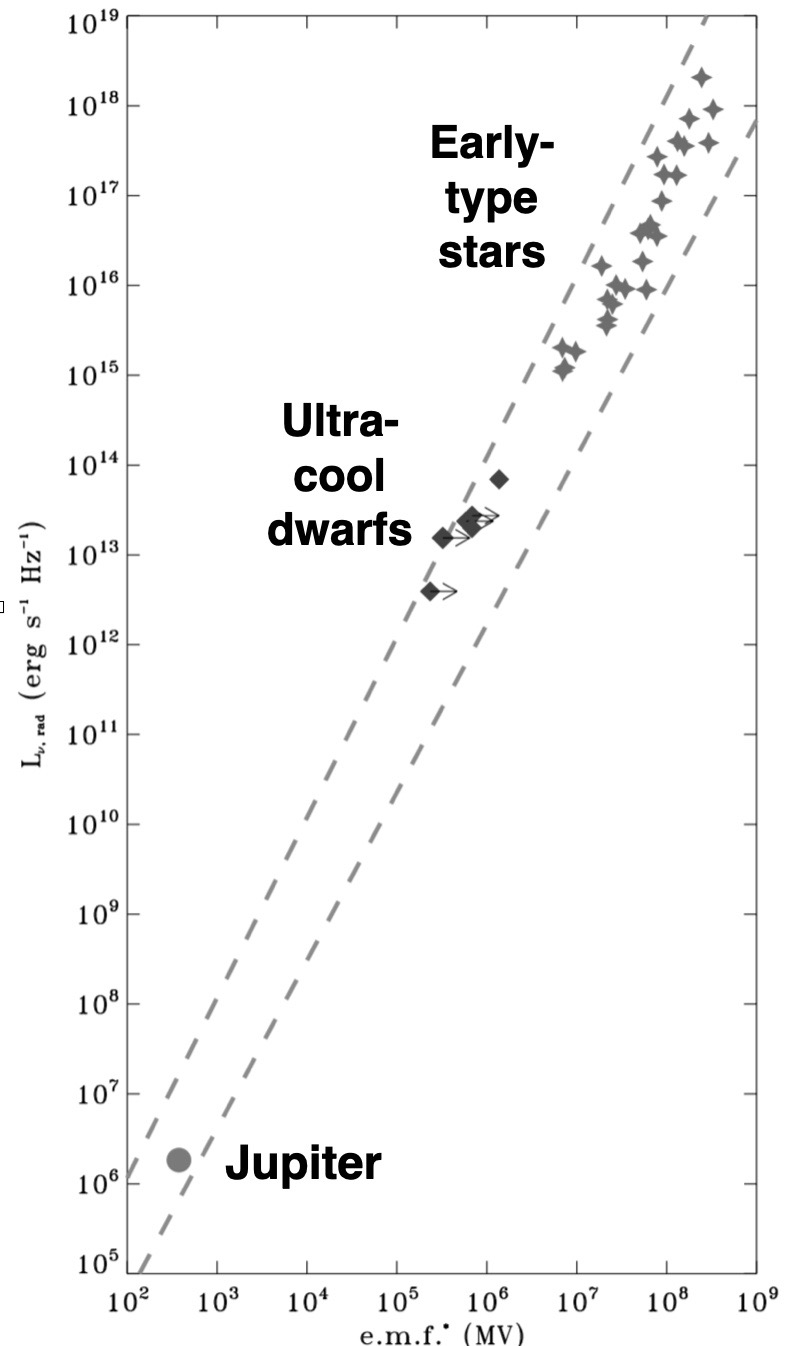}
\vspace{-0.1in} 
 \caption{
 Left: Comparison of observed non-thermal radio luminosity with the luminosity
 scaling for CBO model, showing a strong correlation with an efficiency factor $\epsilon \approx 10^{-8}$
 for conversion to radio emission
 \citep{Owocki22}.
 Right: Empirical scaling law for radio emission, comparing early-type stars with ultra-cool dwarfs, and even
 extending to radio emission from Jupiter \citep{Leto21}.
 }
 \label{fig14} 
\end{center}
\end{figure}

\subsubsection{Role of centrifugal breakout in H$\alpha$ and radio emission}

Despite this success of the RRM model in characterizing the {\it relative}
density of wind-fed material along potential minima, the lack of a clear 
mechanism for emptying this build-up had precluded any prediction of
the {\em absolute} density and associated emission.
But recent theoretical analyses and  numerical MHD simulations
\citep{Owocki20} have provided strong evidence that this wind-fed density 
build-up in CM's proceeds up to a level wherein the outward
centrifugal force exceeds the inward confinement of magnetic,
whereupon material escapes in centrifugal breakout (CBO) 
events.
 This CBO analysis gives a specific, threshold value of the magnetic field strength
 at the Kepler radius, $B_{K1}$, that is needed to give a sufficient density
 for H$\alpha$ to become optically thick, and thus produce a net
 emission within the H$\alpha$  line profile.
 The left panel of figure \ref{fig13} shows a log-log plot
 of  inferred ratio of $B_K/B_{K1}$ vs.  stellar luminosity.
 Remarkably, the onset of H$\alpha$ emission, shown by the filled circles,
 occurs right at the line for unit ratio, with no apparent luminosity 
 trend for the onset of emission.
 The provides strong evidence that CBO is the mechanism for controlling
 density and mass balance in the CM.
 
 Moreover, the right panel of figure \ref{fig13} shows a similar strong 
 dependence on rotation for the observed incoherent {\em radio} emission from such 
 magnetic B-stars \citep{Shultz22}.
 As illustrated in the left side of figure \ref{fig09}, the previously favored model by \citet{Trigilio04}
 proposed that electrons accelerated at the base of the wind-induced current sheet could be 
 spiral along closed magnetic loops, producing the radio  by gyro-synchrotron emission.
 Since this includes no role for stellar rotation, this model is now strongly disfavored.
 
 A promising alternative, illustrated by the right side of figure \ref{fig09}, proposes
 instead that electrons are accelerated near sites magnetic reconnection associated with 
 CBO events.
 An analysis by \citet{Owocki22} show that, under the assumption that the Alfv\'{e}n radius
 follows the split-monopole scaling $R_A \sim \sqrt{\estar} $, the luminosity available
 from such CBO events scales as
 \beq
 L_{CBO} \approx W \Omega B_\ast^2 R_\ast^3
 \, .
 \label{eq:LCBO}
 \eeq
 While this seems to suggest the full volume energy of the magnetic field is dissipated on a rotation timescale,
 in practice the field and its breakout effectively just act as a {\em conduit} for the large energy reservoir associated with the stellar rotation.
 
 The left panel of figure \ref{fig14} compares the observed radio luminosity $L_{rad}$ against this predicted
$L_{CBO}$ scaling \citep{Owocki22}.
 The results show a nearly linear correlation, with however a small efficiency factor $\sim 10^{-8}$ for conversion
 of available CBO energy into radio.
 Moreover, when extended downward by several decades, the right panel shows  that this empirical scaling for radio luminosity
  from early-type stars also matches well that observed from ultra-cool dwarfs \citep{Leto21}.
  Such UCDs are also inferred to have both strong magnetic fields and rapid rotation, and thus are also strong candidates for
  CBO events and associated reconnection.
  
  Even more remarkably,  further downward extrapolation shows that the radio emission from Jupiter follows this same
  scaling relation, thus suggesting a potential link in physical process producing radio emission in stellar and planetary magnetospheres.
  
\begin{figure}[t!]
    \begin{center}
\includegraphics[width=60mm]{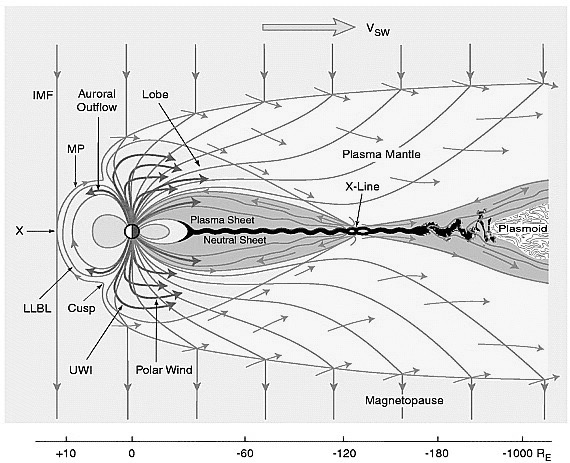}
\includegraphics[width=70mm]{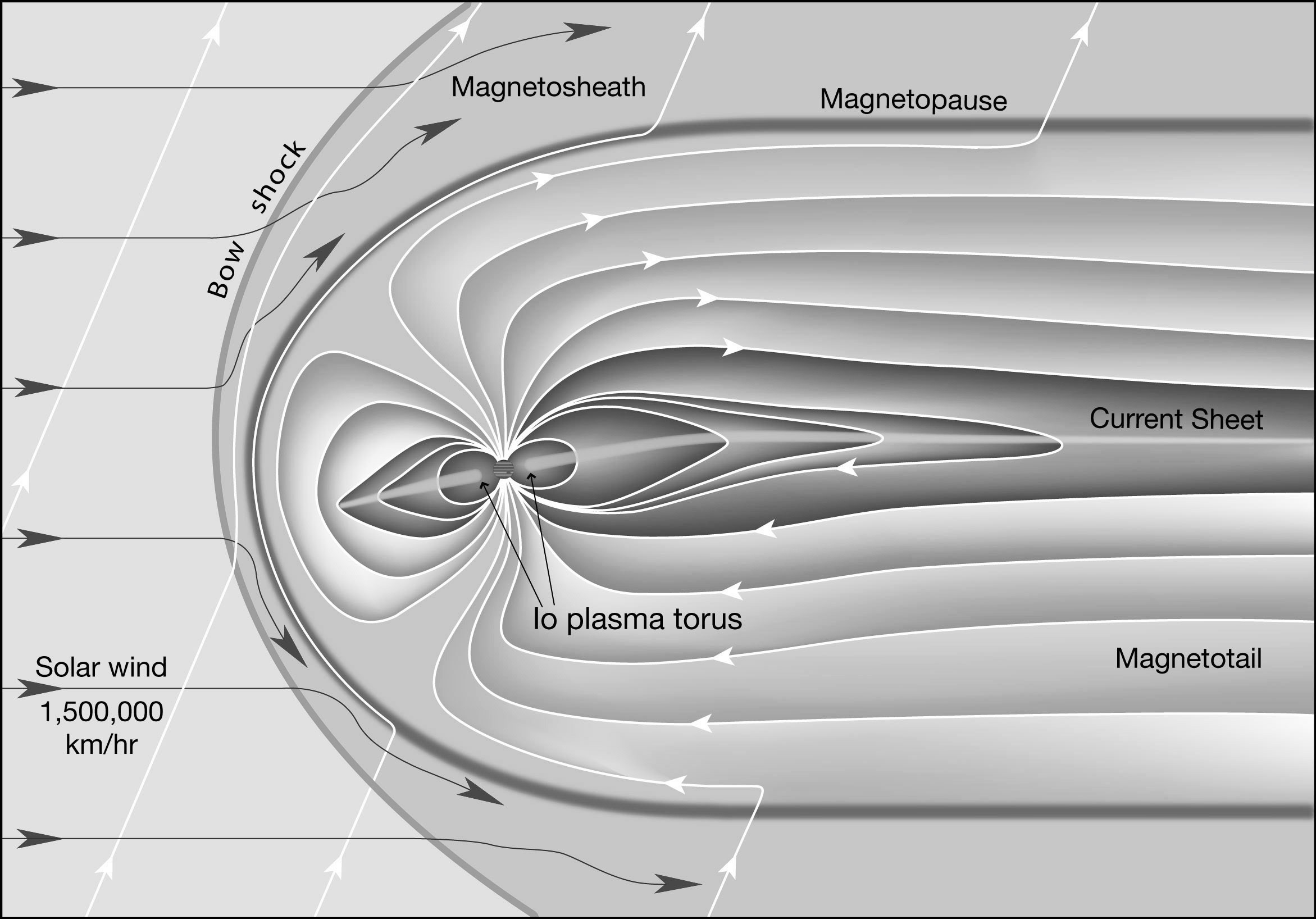}
 \caption{
 Schematics sketches for the magnetospheres of Earth (left) and Jupiter (right).
 Both are externally stressed by the solar wind into elongated tail, leading to episodes
 of magnetotail reconnection that drive particle acceleration manifest as auroral substorms. 
 But Jupiter is also stressed internally from volcanic ejection of material from the active moon Io, 
 which upon ionization by solar UV becomes trapped in Jupiters rotating magnetosphere.
 Jupiter's decametric radio emission is thought link to this Io plasma torus, possibly through
 CBO-driven reconnection and gyro-emission from associated energized electrons.
 Image credits: NASA}
 \label{fig15} 
\end{center}
\end{figure}

\subsection{Outside vs. inside stressing of planetary magnetospheres}

In contrast to the inside-out stressing of stellar magnetospheres, figure \ref{fig15}
illustrates how the impact of an external solar or stellar wind exerts an outside-in stress
on planetary magnetospheres, stretching them into a long magnetotail away from the star.
As seen for the Earth's magnetosphere on the left, thermal expansion now forms a ``polar wind" outflow, which is
then  channeled along  field lines toward the tail.
The closed-field region near the equator traps material into a plasma sheet that extends beyond the Earth's co-rotation radius
($R_K \approx 6 R_E$), where the combination of centrifugal and external stresses from the solar wind drive X-line magnetic 
reconnection and associated plasma ejection along the magnetotail, a process quite analogous to CBO events from stellar magnetospheres.

In the case (right panel) of the much larger magnetosphere of Jupiter, the volcanic moon Io provides an additional plasma source,
at an orbital radius well beyond Jupiter's co-rotation radius, $r_{Io} \approx  2.6 R_K$.
This can combine with other plasmas to drive CBO reconnection, accelerating electrons whose gyro-emission in magnetic loops produce
the decametric radio emission thought linked to Io, again a process similar to the CBO-linked radio emission from stellar magnetospheres.

But a key overriding effect of such planetary magnetospheres is to {\em shield} the planet  from the {\em atmospheric ablation} that
can result from direct impact of coronal wind ions.
Such ablation by the solar wind is thought to be a key factor in depleting the initially dense atmosphere of Mars, which lacks a strong 
global magnetic field.
The presence or absence of sufficiently strong, global magnetic field can thus be a key factor in whether rocky exoplanets retain a sufficient 
atmosphere to host liquid water and perhaps life.

\section{Summary}

Let us conclude with a list of key points of this review:
\begin{itemize}
\item{}~  Hot-star winds are driven by line-scattering of  the star's radiation.
\item{}~   Gas pressure drives supersonic solar coronal winds.
\item{}~   XUV heating of planetary atmospheres drives analogous thermal expansion/escape.
\item{}~   This and ablation deplete atmospheres of rocky planets, but not gas giants.
\item{}~   Solar/stellar magnetospheres guide and trap their wind outflow.
\item{}~   The associated magnetized winds lead to spindown of stellar rotation.
\item{}~   10\% of OBA stars have  strong fields with wind-fed magnetospheres...
\begin{itemize}[label=--]
\item{}~ characterized by the Alfv\'{e}n radius $\ra$ and Kepler co-rotation radius $\rk$.
\item{}~ If $\ra < \rk$, forms a {\em dyanmical magnetosphere} (DM).
\item{}~ If $\rk  < \ra$, forms a {\em centrifugal magnetosphere} (CM).
\end{itemize} 
\item{}~   For CM's, centrifugal breakout (CBO) controls H$\alpha$ and radio emission.
\item{}~   Planetary magnetospheres both shield atmospheres from winds, and  trap outflows.
\end{itemize}

\section*{Acknowledgments}
I thank the IAUS370 SOC, and particularly its chair, Aline Vidotto, for the invitation to give this opening review.
I acknowledge numerous helpful conversations with B. Das, M. Shultz, J. Sundqvist, R. Townsend, A. ud-Doula, and other members
of the Magnetism in Massive Stars (MiMeS) collaboration.
This work was supported in part by NASA ATP grant 80NSSC22K0628.


%
%
%
%

\end{document}